
%
%



\hsize = 6in
\vsize = 8.5in 
\hoffset=0.4cm
\voffset=0.4cm

\catcode`@=11

\font\twelverm=cmr12   
\font\ninerm=cmr9
\font\sevenrm=cmr7

\font\twelvei=cmmi12   
\font\ninei=cmmi9
\font\seveni=cmmi7

\font\twelvesy=cmsy10 scaled 1200  
\font\ninesy=cmsy9
\font\sevensy=cmsy7

\font\tenex=cmex10     

\font\twelvebf=cmbx12  
\font\ninebf=cmbx9
\font\sevenbf=cmbx7

\font\twelvett=cmtt12   

\font\twelvesl=cmsl12   

\font\twelveit=cmti12   

\font\twelvebigbf=cmbx12 scaled 1200
\font\twelveBigbf=cmbx12 scaled 1440

\skewchar\twelvei='177 \skewchar\ninei='177 \skewchar\seveni='177
\skewchar\twelvesy='60 \skewchar\ninesy='60 \skewchar\sevensy='60

\def\twelvepoint{\def\rm{\fam0\twelverm}
          \def\mit{\fam1}
          \def\oldstyle{\fam1\twelvei}
          \def\cal{\fam2}
          \def\it{\fam\itfam\twelveit}
          \def\sl{\fam\slfam\twelvesl}
          \def\bf{\fam\bffam\twelvebf}
          \def\tt{\fam\ttfam\twelvett}
          \def\mbf{\fam\mbffam\twelvembf}
          \def\caps{\fam\capsfam\twelvecaps}
          \def\bigbf{\twelvebigbf}
          \def\Bigbf{\twelveBigbf}
  \textfont0=\twelverm  \scriptfont0=\ninerm  \scriptscriptfont0=\sevenrm
  \textfont1=\twelvei   \scriptfont1=\ninei   \scriptscriptfont1=\seveni
  \textfont2=\twelvesy  \scriptfont2=\ninesy  \scriptscriptfont2=\sevensy
  \textfont3=\tenex  \scriptfont3=\tenex      \scriptscriptfont3=\tenex
  \textfont4=\twelveit
  \textfont5=\twelvesl
  \textfont6=\twelvebf  \scriptfont6=\ninebf  \scriptscriptfont6=\sevenbf
  \textfont7=\twelvett
  \parindent=30pt
  \footindentamount=\parindent
  \parskip=0pt plus 2pt
\def\singlespaces{\abovedisplayskip=11pt plus 3pt minus 3pt
                \belowdisplayskip=11pt plus 3pt minus 3pt
                \abovedisplayshortskip=0pt plus 3pt
                \belowdisplayshortskip=5pt plus 2pt minus 2pt
                \smallskipamount=3pt plus 1pt minus 1pt
                \medskipamount=5pt plus 2pt minus 2pt
                \bigskipamount=11pt plus 4pt minus 4pt
                \alignskipamount=1pt plus1pt minus1pt
                \refbetweenskipamount=7pt plus 2pt minus 1pt
                \normalbaselineskip=14pt
                \normallineskip=1pt
                \parskip=0pt plus 2pt
                \jot=2pt
                \chaplineskipamount=0.15cm
                \sectlineskipamount=3pt
                \chapaboveskipamount=1.0cm plus1.0cm
                \chapbelowskipamount=0.5cm plus0.2cm
                \sectfirstskipamount=0.07cm
                \sectaboveskipamount=0.5cm plus 0.8cm
                \sectbelowskipamount=0.3cm plus 0.1cm
                \normalbaselines}
\def\normalspaces{\abovedisplayskip=15pt plus 4pt minus 4pt
                  \belowdisplayskip=15pt plus 4pt minus 4pt
                  \abovedisplayshortskip=0pt plus 4pt
                  \belowdisplayshortskip=8pt plus 3pt minus 3pt
                  \smallskipamount=3pt plus 1pt minus 1pt
                  \medskipamount=7pt plus 2pt minus 2pt
                  \bigskipamount=15pt plus 5pt minus 5pt
                  \alignskipamount=7pt plus1pt minus1pt
                  \refbetweenskipamount=8pt plus 2pt minus 1pt
                  \normalbaselineskip=18pt
                  \normallineskip=1pt
                  \parskip=0pt plus 2pt
                  \jot=3pt
                  \chaplineskipamount=0.2cm
                  \sectlineskipamount=3pt
                  \chapaboveskipamount=1.5cm plus1.5cm minus0.3cm
                  \chapbelowskipamount=0.3cm plus0.1cm
                  \sectfirstskipamount=0.1cm
                  \sectaboveskipamount=0.7cm plus 1.0cm
                  \sectbelowskipamount=0.3cm plus 0.1cm
                  \normalbaselines}
\def\doublespaces{\abovedisplayskip=22pt plus 6pt minus 6pt
                \belowdisplayskip=22pt plus 6pt minus 6pt
                \abovedisplayshortskip=0pt plus 5pt
                \belowdisplayshortskip=10pt plus 4pt minus 4pt
                \smallskipamount=5pt plus 1pt minus 1pt
                \medskipamount=10pt plus 3pt minus 3pt
                \bigskipamount=22pt plus 7pt minus 7pt
                \alignskipamount=2pt plus1pt minus1pt
                \refbetweenskipamount=14pt plus 3pt minus 1pt
                \normalbaselineskip=24pt
                \normallineskip=1pt
                \parskip=0pt plus 5pt
                \jot=4pt
                \chaplineskipamount=0.3cm
                \sectlineskipamount=5pt
                \chapaboveskipamount=2.0cm plus2.0cm
                \chapbelowskipamount=0.5cm plus0.2cm
                \sectfirstskipamount=0.15cm
                \sectaboveskipamount=1.0cm plus 1.5cm
                \sectbelowskipamount=0.5cm plus 0.2cm
                \normalbaselines}
  \setbox\strutbox=\hbox{\vrule height10.2pt depth4.2pt width0pt}
  \normalspaces\rm}

\newskip\newlineskipamount
\newskip\refbetweenskipamount
\newskip\alignskipamount

\newcount\CHAPNUM  \CHAPNUM=0
\newcount\SECTNUM  \SECTNUM=0
\newcount\REFNUM   \REFNUM=0
\newcount\EQNUM    \EQNUM=0
\newcount\APPNUM   \APPNUM=64
\newcount\EQUATIONCODE \EQUATIONCODE=0
\newcount\SHOWCODE     \SHOWCODE=0

\newskip\newlineskip
\newskip\chaplineskipamount
\newskip\sectlineskipamount
\newskip\chapaboveskipamount
\newskip\chapbelowskipamount
\newskip\sectaboveskipamount
\newskip\sectfirstskipamount
\newskip\sectbelowskipamount
\newskip\footindentamount

\newbox\REFBOX
\newbox\REFDUMPBOX

\def\vfootnote#1{\insert\footins\bgroup
  \parindent=0pt
  \interlinepenalty\interfootnotelinepenalty
  \splittopskip\ht\strutbox 
  \splitmaxdepth\dp\strutbox \floatingpenalty\@MM
  \leftskip\footindentamount \rightskip\z@skip
  \spaceskip\z@skip \xspaceskip\z@skip
  \vskip\medskipamount
  \textindent{\rm #1}\footstrut\rm\futurelet\next\fo@t}

\def\pagenumbers{\global
\footline={\hss\lower0.5cm\hbox{\twelverm\folio}\hss}}
\def\nofirstpagenumber{\global
\footline={\hss\lower0.5cm\hbox{\ifnum\pageno=1{}\else
\twelverm\folio\fi}\hss}}
\pagenumbers


\def\keV{{\rm ke\hskip-0.1em V}}
\def\MeV{{\rm Me\hskip-0.1em V}}

\def\lapprox{\mathrel{\mathop
  {\hbox{\lower0.5ex\hbox{$\sim$}\kern-0.8em\lower-0.7ex\hbox{$<$}}}}}
\def\gapprox{\mathrel{\mathop
  {\hbox{\lower0.5ex\hbox{$\sim$}\kern-0.8em\lower-0.7ex\hbox{$>$}}}}}
\def\eg{{\it e.g.}}
\def\etc{{\it etc.}}
\def\ie{{\it i.e.}}
\def\vs{{\it vs.}}
\def\etal{{\it et~al.}}

\def\pn#1E#2 #3;{#1{\times}10^{#2}\,{\rm#3}}
\def\ex#1;{10^{#1}}
\def\mn#1E#2;{#1{\times}10^{#2}}

\twelvepoint
\def\alignskip{\noalign{\vskip\alignskipamount}}
\def\creturn{\par\nobreak\vskip\newlineskip
   \nobreak\hfil}

\def\title#1{{\parindent=0pt\parskip=0pt
  \newlineskip=0.3cm
  \line{\hfil}
  \vskip1cm plus2cm minus 0.2cm
  \hfil\bigbf#1\par}}
\def\author#1#2{\vskip1.0cm plus 0.3cm minus 0.3cm
  {\newlineskip=0pt\parindent=0pt\parskip=0pt
   \centerline{\rm #1}
   \vskip0.3cm
   \hfil\sl #2\par}}
\def\andauthor#1#2{\vskip0.6cm plus0.2cm minus0.1cm
  \centerline{\rm and}\vskip0.6cm plus0.2cm minus0.1cm
  {\newlineskip=0pt\parindent=0pt\parskip=0pt
   \centerline{\rm #1}
   \vskip0.3cm
   \hfil\sl #2\par}}
\long\def\abstract#1{\vskip1.5cm plus 0.7cm minus 0.4cm
  \centerline{ABSTRACT}
  \nobreak\vskip\chapbelowskipamount\nobreak
  \noindent#1\par}
\def\newpage{\vfil\eject}

\def\chapter#1{{\parindent=0pt\parskip=0pt
  \newlineskip=\chaplineskipamount
  \vskip\chapaboveskipamount\penalty-400
  \global\advance\CHAPNUM by1 \global\SECTNUM=96
  \ifnum\EQUATIONCODE=1\global\EQNUM=0\fi
  \hfil\uppercase\expandafter{\romannumeral\the\CHAPNUM.~#1}\par}
  \nobreak\vskip\chapbelowskipamount\nobreak}
\def\section#1{{\parindent=0pt\parskip=0pt
  \global\advance\SECTNUM by1
  \ifnum\SECTNUM=97\vskip\sectfirstskipamount\nobreak\else
  \vskip\sectaboveskipamount\penalty-300\fi
  \newlineskip=\sectlineskipamount
  \hfil{\it\char\the\SECTNUM) #1}\par}
  \nobreak\vskip\sectbelowskipamount\nobreak}
\def\nochap#1{{\parindent=0pt\parskip=0pt
  \newlineskip=\chaplineskipamount
  \vskip\chapaboveskipamount\penalty-400
  \global\SECTNUM=96
  \hfil\uppercase\expandafter{#1}\par}
  \nobreak\vskip\chapbelowskipamount\nobreak}
\def\appendix#1{{\parindent=0pt\parskip=0pt
  \newlineskip=\chaplineskipamount
  \vskip\chapaboveskipamount\penalty-400
  \global\advance\APPNUM by1 \global\SECTNUM=96
  \global\EQUATIONCODE=2\global\EQNUM=0
  \hfil\uppercase\expandafter{APPENDIX~\char\the\APPNUM:\quad#1}\par}
  \nobreak\vskip\chapbelowskipamount\nobreak}

\def\numberbychapter{\EQUATIONCODE=1}

\def\eq{\global\advance\EQNUM by1\eqno(\the\EQNUM)}
\def\EQ#1{\global\advance\EQNUM by1
  \ifnum\EQUATIONCODE=0 \xdef#1{\the\EQNUM} \fi
  \ifnum\EQUATIONCODE=1 \xdef#1{\the\CHAPNUM.\the\EQNUM} \fi
  \ifnum\EQUATIONCODE=2 \xdef#1{{\rm\char\the\APPNUM.\the\EQNUM}}\fi
  \ifnum\SHOWCODE=0 \eqno(#1) \fi
  \ifnum\SHOWCODE=1 \eqno(#1)\hbox to0pt{\tt\ \string#1\hss}\fi}
\def\frac#1#2{{#1\over#2}}
\def\half{{\textstyle{1\over2}}}
\def\quarter{{\textstyle{1\over4}}}

\catcode`@=12 

\setbox\REFBOX=\vbox{\nochap{References}}
\def\RF#1#2{\gdef#1{\global\advance\REFNUM by1
  \global\setbox\REFBOX=\vbox{\unvbox\REFBOX
    \parindent=25pt
    \vskip\refbetweenskipamount\goodbreak
    {\leftskip=0pt\rightskip=0pt\rm\item{\the\REFNUM.}#2\par}}
    \the\REFNUM\xdef#1{\the\REFNUM}}}
\def\qref#1{\unskip$^{#1}$}
\def\refout{\unvbox\REFBOX}
\def\dumpref#1{\setbox\REFDUMPBOX=\vbox{#1}-}
\def\journal#1; #2;{\gdef#1##1 (##2) ##3;{{\sl#2\ }{\bf##1},
    {\rm##3 (##2)}}}
\def\Journal#1; #2; #3;{\gdef#1##1 (##2) ##3;{{\sl#2\ }%
{\bf##1\kern0.2em#3}, {\rm##3 (##2)}}}
\def\JOURNAL#1; #2; #3;{\gdef#1##1 (##2) ##3;{{\sl#2\ }%
{\bf#3\kern0.2em##1}, {\rm##3 (##2)}}}

\journal\Ib; {\it ibid.};

\journal\AA; Astron.\ Astrophys.;
\journal\AAR; Astron.\ Astrophys.\ Rev.;
\journal\AJ; Astron.~J.;
\journal\AJP; Austr.~J.\ Phys.;
\journal\AN; Astr.\ Nachr.;
\journal\ANP; Adv.\ Nucl.\ Phys.;
\journal\ANYAS; Ann.\ N.~Y.\ Acad.\ Sci.;
\journal\ApJ; Astrophys.~J.;
\journal\ApJS; Astrophys.~J.\ Suppl.;
\journal\ApL; Astrophys.\ Lett.;
\journal\APNY; Ann.\ Phys.\ (N.Y.);
\journal\APP; Astropart.\ Phys.;
\journal\ARAA; Ann.\ Rev.\ Astron.\ Astrophys.;
\journal\ARNS; Ann.\ Rev.\ Nucl.\ Sci.;
\journal\ARNPS; Ann.\ Rev.\ Nucl.\ Part.\ Sci.;
\journal\ASS; Ap.\ Sp.\ Sci.;
\journal\CA; Comm.\ Astrophys.;
\journal\CNPP; Comm.\ Nucl.\ Part.\ Phys.;
\journal\CQG; Class.\ Quantum Grav.;
\journal\CS; Curr.\ Sci.;
\journal\DAN; Dokl.\ Akad.\ Nauk.\ S.S.S.R.;
\journal\FCP; Fund.\ Cosmic Phys.;
\journal\FP; Fortschr.\ Phys.;
\journal\HPA; Helv.\ Phys.\ Acta;
\journal\IJTP; Indian J.\ Theor.\ Phys.;
\journal\JETP; Sov.\ Phys.\ JETP;
\journal\JETPL; JETP Lett.;
\journal\MNRAS; Mon.\ Not.\ R.\ Astron.\ Soc.;
\journal\MSAI; Mem.\ Soc.\ Astron.\ Ital.;
\journal\MPLA; Mod.\ Phys.\ Lett.~A;
\journal\Nature; Nature;
\journal\NC; Nuovo Cim.;
\Journal\NCA; Nuovo Cim.; A;
\Journal\NCC; Nuovo Cim.; C;
\journal\NCS; Nuovo Cim.\ Suppl.;
\journal\NIM; Nucl.\ Instr.\ and Meth.;
\JOURNAL\NPA; Nucl.\ Phys.; A;
\JOURNAL\NPB; Nucl.\ Phys.; B;
\journal\PASJ; Publ.\ Astron.\ Soc.\ Japan;
\journal\PAZ; Pisma Astr.\ Zh.;
\Journal\PLA; Phys.\ Lett.; A;
\Journal\PLB; Phys.\ Lett.; B;
\journal\PPNP; Prog.\ Part.\ Nucl.\ Phys.;
\journal\PTRSLA; Phil.\ Trans.\ Roy.\ Soc.\ London~A;
\journal\PRep; Phys.\ Rep.;
\journal\PR; Phys.\ Rev.;
\journal\PRL; Phys.\ Rev.\ Lett.;
\journal\PRA; Phys.\ Rev.\ A;
\journal\PRB; Phys.\ Rev.\ B;
\journal\PRC; Phys.\ Rev.\ C;
\journal\PRD; Phys.\ Rev.\ D;
\JOURNAL\PRSA; Proc.\ Roy.\ Soc.\ Lond.; A;
\journal\PTP; Prog.\ Theor.\ Phys.;
\journal\PZ; Phys.~Z.;
\journal\PZETF; Pisma Zh.\ Eksp.\ Teor.\ Fiz.;
\journal\RMP; Rev.\ Mod.\ Phys.;
\journal\RNC; Riv.\ Nuovo Cim.;
\journal\RPP; Rep.\ Prog.\ Phys.;
\journal\SA; Sci.\ Am.;
\journal\SAL; Sov.\ Astron.\ Lett.;
\journal\Science; Science;
\journal\SJNP; Sov.\ J.\ Nucl.\ Phys.;
\journal\SP; Solar Phys.;
\journal\SST; Speculations\ Sci.\ Tech.;
\journal\YF; Yad.\ Fiz.;
\journal\ZP; Z.~Phys.;
\journal\ZPC; Z.~Phys.\ C;
\journal\ZETF; Zh.\ Eksp.\ Teor.\ Fiz.;

\def\qref#1{$[#1]$}
\numberbychapter
\nofirstpagenumber
\singlespaces
\hyphenation{brems-strah-lung}

\def\eref#1{Eq.~(#1)}

\def\SS{{\cal S}}
\def\A{{\bf A}}
\def\x{{\bf x}}
\def\k{{\bf k}}
\def\p{{\bf p}}

\def\NN{{N\!N}}

\def\GF{G_{\rm F}}
\def\nubar{{\overline\nu}}
\def\o{\omega}
\def\otil{\widetilde\omega}

\def\flip{{\rm flip}}
\def\pair{{\rm pair}}
\def\scat{{\rm scat}}

\def\gamscat{\Gamma_{\rm scat}}
\def\gampair{\Gamma_{\rm pair}}

\def\fscat{F_{\rm scat}}
\def\fpair{F_{\rm pair}}
\def\norm{\frac{\GF^2 N_B}{4\pi^4 N_\nu}}

\def\gamint{\Gamma_{\rm int}}
\def\omin{\omega_{\rm min}}

\def\ka{k_a}

\def\kpi{k_\pi}
\def\kkpi{\k_\pi}
\def\kpo{\k_{\pi0}}
\def\opi{\o_\pi}
\def\opo{\o_{\pi0}}
\def\mupi{\mu_\pi}
\def\SPI{S_\pi^{\mu\nu}}

\def\caa{C_{A,0}}
\def\cab{C_{A,1}}
\def\cva{C_{V,0}}
\def\abs#1{\vert#1\vert}

\def\fF{1}
\def\fGamma{2}
\def\fQ{3}



\RF\BurrowsGT{A.~Burrows, R.~Gandhi, and M.~S.~Turner,
    \PRL 68 (1992) 3834;.}
\RF\GaemersGL{K.~J.~F.~Gaemers, R.~Gandhi, and J.~M.~Lattimer,
    \PRD 40 (1989) 309;.}
\RF\GandhiB{R.~Gandhi and A.~Burrows, \PLB 246 (1990) 149; and
    \Ib 261 (1991) 519(E);.
    J.~A.~Grifols and E.~Mass\'o, \PLB 242 (1990) 77;.
    A.~P\'erez and R.~Gandhi, \PRD 41 (1990) 2374;.
    A.~A.~Natale, \PLB 258 (1991) 227;.}
\RF\Turner{M.~Turner, \PRD 45 (1992) 1066;.}
\RF\LamN{W.~P.~Lam and K.-W.~Ng, \PLB 284 (1992) 331;.}
\RF\Pantaleone{J.~Pantaleone, \PLB 268 (1991) 227;.}
\RF\RaffeltS{G.~Raffelt and D.~Seckel, \PRL 60 (1988) 1793;.}
\RF\Maalampi{J.~Maalampi and J.~T.~Peltoniemi, \PLB 269 (1991) 357;.
    G.~Raffelt and G.~Sigl, \APP 1 (1993) 165;.}
\RF\Wilson{R.~Mayle, D.~N.~Schramm, M.~S.~Turner, and J.~R.~Wilson,
    \PLB 317 (1993) 119;.}

\RF\SawyerI{R.~F.~Sawyer, \PRD 11 (1975) 2740;.}
\RF\SawyerII{R.~F.~Sawyer, \PRL 61 (1988) 2171; and \PRC 40 (1989) 865;.}
\RF\SawyerS{R.~F.~Sawyer and A.~Soni, \ApJ 230 (1979) 859;.}
\RF\IwamotoP{N.~Iwamoto and C.~J.~Pethick, \PRD 25 (1982) 313;.}

\RF\FrimanM{B.~L.~Friman and O.~V.~Maxwell, \ApJ 232 (1979) 541;.}

\RF\Glendenning{See, for example, K.~Glendenning, \NPA 493 (1989) 521;.}
\RF\HaenselJ{P.~Haensel and  A.~J.~Jerzak, \AA 179 (1987) 127;.}
\RF\HorowitzW{C.~J.~Horowitz and K.~Wehrberger, \PLB 266 (1991) 236;
    and \PRL 66 (1991) 272;.}
\RF\HorowitzS{C.~J.~Horowitz and B.~D.~Serot, \NPA 464 (1987) 613;.}
\RF\Kirzhnits{D.~A.~Kirzhnits, V.~V.~Losyakov, and V.~A.~Chechin,
    \ZETF 97 (1990) 1089; [\JETP 70 (1990) 609;].}

\RF\RaffeltSS{G.~Raffelt and D.~Seckel, \PRL 67 (1991) 2605; and
    \PRL 68 (1992) 3116;.
    R.~Sawyer, \PRL 68 (1992) 3115;.}
\RF\Sigl{See, for example, G.~Raffelt, G.~Sigl, and L.~Stodolsky,
    \PRL 70 (1993) 2363;; G.~Sigl and G.~Raffelt, \NPB 406 (1993) 423;.}
\RF\Pioncondensate{J.~N.~Bahcall and R.~A.~Wolf, \PR 140 (1965) B~1452;.
     G.~Baym, \PRL 30 (1973) 1340;.
     O.~Maxwell, G.~E.~Brown, D.~K.~Cambell, R.~F.~Dashen, and
     J.~T.~Manassah, \ApJ 216 (1977) 77;.
     For a recent review see A.~B.~Migdal \etal, \PRep 192 (1990) 179;.}

\RF\SNReview{For reviews see D.~N.~Schramm, \CNPP 17 (1987) 239;;\hfil\break
    D.~N.~Schramm and J.~W.~Truran, \PRep 189 (1990) 89;.}
\RF\BurrowsL{A.~Burrows and J.~L.~Lattimer, \ApJ 307 (1986) 178;.}
\RF\Seckel{D.~Seckel and G.~Raffelt, work in progress (1993).}

\RF\Raffelt{G.~G.~Raffelt, \MPLA 5 (1990) 2581;.}
\RF\AxReview{For reviews see M.~S.~Turner, \PRep 197 (1990) 67;;
    G.~G.~Raffelt, \PRep 198 (1990) 1;.}
\RF\Weldon{A.~Weldon, \PRD 28 (1983) 2007;.}

\RF\axnucl{See, for example, B.~Castle and I.~Towner,
 {\it Modern Theories
 of Nuclear Moments} (Clarendon Press, Oxford, 1990). }
\RF\EMCa{EMC Collaboration, J.~Ashman, \etal, \PLB 206 (1988) 364;;
  \NPB 328 (1989) 1;.}
\RF\SMC{SMC Collaboration, B.~Adeva, \etal, \PLB 302 (1993) 553;.}
\RF\Eslac{E142 Collaboration, P.L.~Anthony, \etal, \PRL 71 (1993) 959;.}

\RF\EllisK{J. Ellis and M.~Karliner, CERN-TH.7022/93 (Talk presented at
13th International Conference on Particles and Nuclei, PANIC '93, Perugia,
Italy).}


\line{December 1993\hfil MPI-Ph/93-90}
\line{\hfil BA-93-43}
\title{A Self-Consistent Approach to Neutral-Current Processes\creturn
in Supernova Cores}

\author{Georg~Raffelt}
       {Max-Planck-Institut f\"ur Physik\creturn
       F\"ohringer Ring 6, 80805 M\"unchen, Germany}
       \andauthor{David~Seckel}
       {Bartol Research Institute, University of Delaware\creturn
       Newark, DE 19716, U.S.A.}

\abstract{The problem of neutral-current processes (neutrino scattering,
pair emission, pair absorption, axion emission, \etc)
in a nuclear medium can be separated into an
expression representing the phase space of the weakly interacting probe, and a
set of dynamic structure functions of the medium. For a non-relativistic medium
we reduce the description to two structure functions $S_A(\o)$ and $S_V(\o)$ of
the energy transfer, representing the axial-vector and vector interactions.
$S_V$ is well determined by the single-nucleon approximation while $S_A$ may be
dominated by multiply interacting nucleons.
Unless the shape of $S_A(\o)$ changes dramatically at high densities,
scattering processes always dominate over pair processes
for neutrino transport or the emission of right-handed
states. Because the emission of right-handed neutrinos and axions is
controlled by the same medium response functions, a consistent constraint
on their properties from consideration of supernova cooling should
use the same structure functions for both neutrino transport and exotic cooling
mechanisms.}

\newpage


\chapter{Introduction}

\noindent The neutrino signal from the supernova (SN) 1987A has confirmed
our basic understanding that type~II supernovae result from the core collapse
of massive stars~\qref\SNReview. The observed neutrinos were radiated from the
``neutrino-sphere'' with a luminosity and energy spectrum commensurate with
expectations based both on broad theoretical grounds and detailed numerical
models.  Further, the time scale for the neutrino ``light curve'' was broadly
in agreement with the hypothesis that energy transport in the hot, dense core
is dominated by neutrino diffusion. Despite these successes there is much that
needs to be done to achieve both a quantitative and a qualitative
understanding of SN physics.

In this paper, we examine a number of issues concerning neutrino transport in
SN cores. We are especially interested in the neutral-current (NC) processes
that govern the transport of $\mu$ and $\tau$ neutrinos. These include
scattering $\nu X\to X'\nu$, pair emission $X\to X'\nu\overline\nu$, and pair
absorption $\nu\overline\nu X\to X'$ where $X$ and $X'$ are configurations of
one or several particles of the medium. These processes are intimately related
because the underlying matrix elements are structurally the same.  Still, the
pair processes have usually been neglected in transport
calculations~\qref\BurrowsL, presumably because in a non-interacting gas of
nucleons they vanish due to energy momentum conservation. We would like to
know if this naive expectation is born out in a strongly interacting medium.

Although this question has always been present, it has been obscured by the
difficulty of calculating any of the relevant processes reliably. Further, the
authors that worried about scattering processes and transport in hot SN
cores~\qref{\SawyerI,\SawyerII,\SawyerS,\IwamotoP}, could plausibly assume
that scattering was the dominant part of neutrino transport; whereas authors
worried about pair processes were usually concentrating on the cooling of
older, cold, degenerate neutron stars~\qref\FrimanM. There is another problem
in that the relevant calculations are usually done in the context of
perturbation theory---processes with the fewest number of nucleons are assumed
to dominate. The lowest-order emission process is usually neutrino pair
bremsstrahlung involving two nucleons, $NN \to NN\nu\overline\nu$, whereas the
lowest order scattering process is scattering from a single nucleon
$\nu N \to N\nu$.  Therefore, if pair processes turned out to be important
compared to scattering, the whole perturbative framework of the calculations
would seem questionable.

In this light, we find several recent papers on the SN emission of
hypothetical right-handed (r.h.) neutrinos extremely
interesting~\qref{\Turner,\BurrowsGT,\Wilson}.
On the basis of perturbative calculations of nucleon-nucleon and pion-nucleon
interactions (if there is a pion condensate) it appeared that pair processes
of the type $X\to X'\overline\nu_L\nu_R$ would be more important than
spin-flip scattering $\nu_LX\to X'\nu_R$ in a non-degenerate nuclear medium.
As these processes are structurally very similar to the ones involving only
left-handed (l.h.) neutrinos we are led to wonder if pair processes are also
important for the standard neutrino transport in newborn neutron stars,
and under what, if any, conditions the perturbative framework can be trusted.
Similarly, resolving the question of pair processes in transport calculations
will affect the interpretation of the emission rates calculated for several
species of hypothetical particles, including r.h.~neutrinos and
axions.

General answers to these questions are difficult because of our lack of
knowledge about strongly interacting systems. In the end we travel the
perturbative road, but before doing so we want to make sure that the relation
between scattering and pair processes is kept clear. To this end, we factorize
the problem of neutrino interactions into a ``neutrino part'', and a ``medium
part''. This allows us to maintain a consistent treatment of the medium while
discussing the different neutrino processes. The medium is then described by a
small number of response functions, common to all NC  processes.  In the limit
of non-relativistic medium constituents and ignoring the neutrino momentum
transfer to the nucleons, the medium response can be reduced to a single
structure function $S(\o)$ of the energy transfer.  $S$ is a
linear combination of the density and spin-density dynamic structure
functions at vanishing three-momentum transfer.  Although we derive $S$ for
ordinary l.h.~neutrino interactions it also applies to spin-flip scattering.

In a sufficiently dilute medium the dynamic structure functions can be
calculated perfectly well by perturbative methods, so we may borrow readily
from other authors' calculations. Our hope is that when extended into the
regime of high densities the overall shape of $S(\o)$ will not abruptly
change. In this case we will argue that the {\it relative\/} strength of
scattering almost certainly exceeds that of pair processes for a medium
dominated by non-relativistic nucleons. Thus, if a pion condensate would
substantially add to $X\to X'\nu_R\overline\nu_L$ it would add even more to
$\nu_L X\to X'\nu_R$.  Then, however, it would also strongly affect $\nu_L
X\to X'\nu_L$ and thus neutrino transport.

Turn back to the emission of weakly interacting particles.
The bremsstrahlung emission of r.h.~neutrinos or axions
are sensitive only to the spin-density fluctuations. Even so, the
spin-density structure function used for r.h.~neutrino
or axion emission is the same as
that for ordinary l.h.~neutrino scattering and, therefore, the
uncertainties of, for example, the axion emission rate are not unrelated to
that of neutrino transport. It follows that a consistent treatment of
``exotic''
particle emission would have to rely on a common structure function $S(\o)$
for both neutrino transport and particle emission, whether or not $S(\o)$ can
be reliably calculated.  Modifications of the neutrino transport and the
emission of novel particles will both affect the observable neutrino signal,
and thus they should be implemented on the same footing.

There remains the problem of how to calculate the dynamic structure
functions, a problem which is arguably more difficult for the present
application than for any other problem concerning neutrino interactions with
stellar material. The medium at the core of a proto-neutron star is hot, dense
and strongly interacting. There are no parameters that one may use as the
basis for a perturbation expansion. Further, the material is neither degenerate
nor non-degenerate. In this context, we suggest a phenomenological approach
that we believe illustrates some of the features that a full treatment of an
interacting medium must possess.  This approach uses the interactions of the
medium to regulate soft processes, $\ie$ those where the energy transfer is
small, but leaves the hard processes relatively unscathed. Although not
totally successful, the technique goes in the direction
of regulating the amplitude of the response function calculated in a
perturbative series, while modifying its shape in a controlled way.

The rest of this paper is organized as follows.
In \S~II we develop the main tool for our investigation, the
vector and axial structure functions $S_V(\o)$ and $S_A(\o)$, respectively,
which describe the medium's response to NCs.
In \S~III we present useful properties of these functions and
evaluate them perturbatively.
In \S~IV we examine the relative strength of pair vs.\ scattering processes
under a variety of different assumptions about the high-density behavior
of $S_A$.
In \S~V we consider applications to non-standard physics, \ie,
the emission of novel particles and the impact of a pion condensate.
Finally, in \S~VI, we summarize our discussions and conclusions,
with some suggestions as to useful strategies for future research.



\chapter{Neutral-Current Neutrino Processes in a Medium}
\section{Collision Integral}
\noindent Neutrino transport in a medium is governed by the Boltzmann
collision equation, $L[f]=C[f,\overline f]\equiv(df/dt)_{\rm coll}$
where $f$ and $\overline f$ are the neutrino and anti-neutrino occupation
numbers for l.h.~states. (For now we focus on the standard model
without r.h.~neutrinos.) An analogous equation applies to $\overline
f$. The Liouville operator is given by
$L[f]=\partial_t f+\dot\x{\cdot}\nabla_\x f+\dot\k{\cdot}\nabla_\k f$
while the collision integral is
$$
\eqalign{\left.\frac{df_{\k_1}}{dt}\right\vert_{\rm coll}=
  \int\frac{d^3\k_2}{(2\pi)^3}\,\biggl[&W_{k_2,k_1}f_{\k_2}(1-f_{\k_1})
  -W_{k_1,k_2}f_{\k_1}(1-f_{\k_2})\,+\cr
  &+W_{-k_2,k_1}(1-f_{\k_1})(1-\overline f_{\k_2})
  -W_{k_1,-k_2}f_{\k_1}\overline f_{\k_2}\biggr]\,,\cr}
  \EQ\Wa
$$
where we have written the momentum variables as subscripts. The first term
corresponds to neutrino scatterings into the mode $\k_1$ from all other modes,
the second term is scattering out of mode $\k_1$ into all other modes, the
third term is pair production with a final-state neutrino $\k_1$, and the
fourth term is pair absorption of a neutrino of momentum $\k_1$ and an
anti-neutrino of any momentum. In this collision integral we only include
effective NC processes between neutrinos and the nuclear
medium, \ie, we are not considering charged-current processes and
NC processes coupling neutrinos to the leptons in the
medium.\footnote{$^\dagger$}{The temperature $T$ is expected to be of order
$50\,\MeV$, whereas the density of baryons is characterized by Fermi momenta
in the $400-500\,\MeV$ range. Particle
species whose density is characterized by $T^3$
are not nearly as abundant as nucleons. The importance of degenerate $e^-$ and
$\nu_e$ for neutrino scattering is typically less than a tenth that of the
nucleons. Even though it is conceivable that there may be situations where
they dominate~\qref{\LamN},
we will always ignore purely leptonic processes.}

The expression $W_{k_1,k_2}=W(k_1,k_2)$ is the rate for a neutrino in state
$k_1$ to scatter into $k_2$ via interaction with the medium. $W$ is defined in
the entire $k$ space for both of its arguments where negative energies
indicate the ``crossing'' of an initial-state neutrino into a final-state
anti-neutrino or vice versa.  In thermal equilibrium $f$ and $\overline f$
must assume Fermi-Dirac distributions at a given temperature $T$ and chemical
potential $\mu$ while the collision integral must vanish, leading to the
detailed-balance requirement $W(k_1,k_2)=e^{(\o_1-\o_2)/T}W(k_2,k_1)$.

It is the pair production and absorption terms under the integral in Eq.~(\Wa)
that are usually neglected in the context of SN neutrino transport. The
justification appears to be that the pair terms vanish for a medium of free
nucleons. More specifically, $W(k_1,k_2)$ is identically zero if the energy
momentum transfer $k=k_1-k_2$ is timelike ($k^2 > 0$) and if the dispersion
relations of the medium excitations are like that of ordinary particles,
$E^2-p^2=m^2>0$.

\section{The Transition Rate $W(k_1,k_2)$}
\noindent
The low-energy NC Hamiltonian for the neutrino field $\psi$ and
an effective current operator $B^\mu$ for the medium is
$$
H_{\rm int}=\frac{\GF}{2\sqrt2}\,B^\mu\,
  \overline\psi\gamma_\mu(1-\gamma_5)\psi\,.
  \EQ\Wb
$$
If the neutrino interactions are dominated by nucleons, the current of the
background medium is
$$
  B^\mu = \sum_{i=n,p}
  \overline\psi_i\gamma^\mu(C_{V,i}-C_{A,i}\gamma_5)\psi_i\,,
  \EQ\Wee
$$
where $\psi_{n,p}$ are the interacting quantum fields for protons and
neutrons, and $C_{V,i}$ and $C_{A,i}$ are the relevant vector and axial weak
charges.

The current-current structure of $H_{\rm int}$ allows the transition rate to
be written in the general form
$$
W(k_1,k_2)=\frac{\GF^2 N_B}{8}\,S_{\mu\nu}N^{\mu\nu}\,,\EQ\Wc
$$
where $\GF$ is the Fermi constant and $N_B$ the number density of baryons.
$N^{\mu\nu}$ entails the neutrino kinematics, and $S_{\mu\nu}$ describes
fluctuations in the medium that produce, scatter, or absorb neutrinos, \ie, it
is the dynamic structure function of the medium for NCs.

Up to a normalization factor, $N^{\mu\nu}$ is given by the neutrino part of
the squared matrix element for the scattering $\nu_{\k_1}\to\nu_{\k_2}$
\qref\GaemersGL\
$$
N^{\mu\nu}=\frac{8}{2\o_12\o_2}\,
  \Bigl[k_1^\mu k_2^\nu+k_2^\mu k_1^\nu-k_1k_2\, g^{\mu\nu}
  -i\epsilon^{\alpha\beta\mu\nu}k_{1\alpha}k_{2\beta}\Bigr]\,.
  \EQ\Wd
$$
Then, by the usual crossing relations the expression for the emission of a
pair $\overline\nu_{\k_1}\nu_{\k_2}$ is given by $k_1\to-k_1$, while
$k_2\to-k_2$ is for the absorption of $\nu_{\k_1}\overline\nu_{\k_2}$. Both
operations leave $N^{\mu\nu}$ unchanged because we included a factor
$(2\o_12\o_2)^{-1}$ in its definition.

The dynamic structure factor is then functionally dependent only on the energy
momentum transfer so that a single tensor $S_{\mu\nu}(k)$ describes all
processes in Eq.~(\Wa). It can be written directly in terms of the
fluctuations in the weak NCs~\qref\Sigl,
$$
S^{\mu\nu}(k)=\frac{1}{N_B}\int_{-\infty}^{+\infty} dt\,e^{i\omega t}
  \left\langle B^\mu(t,\k) B^\nu(0,-\k)\right\rangle\,,
  \EQ\We
$$
where $k=k_1-k_2=(\o,\k)$.
By the expectation value $\langle\ldots\rangle$ of some operator we
mean a trace over a thermal ensemble of the background medium.

Equation~(\Wc) is a first-order perturbative result in the weak
Hamiltonian.
The medium, on the other hand, is strongly interacting, so a perturbative
calculation of $S_{\mu\nu}$ may or may not be possible.
Of course, in principle some insight into its properties can be gained by
laboratory experiments, \eg\ particle emission from the hot and dense systems
produced in heavy ion collisions.

However, for practical SN calculations one proceeds with a perturbative
approach as follows. For a particular term in the expansion one calculates the
squared matrix element, which can be written in the form
$\vert{\cal M}\vert^2=(\GF^2/8)\,M_{\mu\nu} N^{\mu\nu} 2\omega_12\omega_2$,
where $M^{\mu\nu}$ is the square of the medium part of the matrix element.
Then one performs a phase-space integral over all medium participants and sums
over all relevant processes,
$$
\eqalign{S_{\mu\nu}=\frac{1}{N_B}
  \sum_{\rm processes}
  \prod\limits_{i=1}^{N_i}\int\frac{d^3\p_i}{2E_i(2\pi)^3}\,f_{\p_i}
  \prod\limits_{f=1}^{N_f}\int\frac{d^3\p_f}{2E_f(2\pi)^3}\,
  (1\pm f_{\p_f})\,\times&\cr
  \alignskip
  \times\,(2\pi)^4\,
  \delta^4\left(k+\sum_{i=1}^{N_i} p_i-\sum_{f=1}^{N_f} p_f \right)
  \sum_{\rm spins}M_{\mu\nu}&\,,}
  \EQ\Wf
$$
where $N_i$ and $N_f$ are the number of initial and final medium particles,
the $f_{\p}$ are occupation numbers, and the $(1\pm f_\p)$ are Pauli-blocking
or Bose-stimulation factors.

Unfortunately, for the conditions of a SN core higher-order processes
typically yield contributions to $S_{\mu\nu}$ which are of the same order or
larger than lower-order ones (see \S~III below), so one may question the
validity of this method. However, because $W(k_1,k_2)$ factorizes, this
complication can be separated from a discussion of the relative importance of
scattering \vs\ pair processes, which is largely an issue of neutrino phase
space.

Therefore, instead of relying too much on perturbative calculations of
$S^{\mu\nu}(k)$ we should focus on its general properties.
To this end we assume that the medium is homogenous and isotropic, in which
case $S^{\mu\nu}$ can be constructed only from the energy-momentum
transfer $k$ and the four-velocity $u$ of the medium~\qref\Kirzhnits,
$$
  S^{\mu\nu}=R_1\,u^\mu u^\nu + R_2\,(u^\mu u^\nu-g^{\mu\nu})
  +R_3\,k^\mu k^\nu +R_4\,(k^\mu u^\nu+u^\mu k^\nu)
  +iR_5\,\epsilon^{\mu\nu\alpha\beta}u_\alpha k_\beta\,.\EQ\Wg
$$
The structure functions $R_1,\ldots,R_5$ depend on the medium temperature and
chemical composition, and on the Lorentz scalars that can be constructed from
$u$ and $k$, namely $k^2$ and $uk$. (The third possibility $u^2=1$ is a
constant.) Instead of $k^2$ and $uk$ we will use as independent variables the
energy and momentum transfer $\o$ and $\vert\k\vert$, measured in the rest
frame of the medium.

It is convenient to calculate the interaction rate in the rest frame of the
medium defined by $u=(1,0,0,0)$. Multiplying Eqs.~(\Wd) and~(\Wg), we find
$$
  W(k_1,k_2)= \frac{\GF^2N_B}{4}
  \biggl[(1+\cos\theta)\,R_1+(3-\cos\theta)\,R_2
  -2(1-\cos\theta)(\o_1+\o_2)\,R_5 \biggr]
  \,,\EQ\Wh
$$
where $\theta$ is the neutrino scattering angle. It is specific to the
contraction with  $N^{\mu\nu}$, relevant for l.h.~neutrinos, that only
$R_{1,2,5}$ contribute. For the spin-flip processes discussed in \S~V we will
find that all $R$'s contribute while for axion emission only $R_2$
survives.

\section{Non-Relativistic and Long-Wavelength Limit}
\noindent
Within the context of a nucleonic medium there are two closely related limits
that are often taken---the ``non-relativistic'' and ``the long-wavelength''
approximations. The justification for both is that the nucleon mass is larger
than any other energy or momentum scale in the problem. (Even though nucleon
Fermi momenta are in the $400-500\,\MeV$ range and the effective nucleon mass
may be as low as $600\,\MeV$, it is still reasonable to treat the nucleons
non-relativistically for the purposes of this paper.) Then, the currents are
expanded as a power series in $1/m_N$, after which only the leading term is
kept.\footnote{$^\dagger$}{It may happen that the leading contribution to the
current makes no contribution to an interaction, in which case
the next term in the non-relativistic expansion should be kept.}
The long-wavelength approximation assumes that the three-momentum transfer to
the nucleons is small compared to typical nucleon momenta and thus can be
ignored when calculating the available phase-space, thus simplifying those
calculations considerably.

As a consequence of the non-relativistic assumption the structure functions
$R_{3,4,5}$ vanish. To see this we first note that the medium current in
Eq.~(\Wb) generally is a sum of a vector and an axial-vector piece,
$B^\mu=V^\mu+A^\mu$. Then the transformation properties under parity of the
five terms in Eq.~(\Wg) are easily identified to be like
$\langle V^\mu V^\nu\rangle$ or $\langle A^\mu A^\nu\rangle$
for terms $1-4$ and $\langle A^\mu V^\nu+V^\mu A^\nu\rangle$ for term~5.
In the limit of non-relativistic nucleons $V^\mu$ has only a zero-component,
$V^0=C_V\psi^\dagger\psi$ (a sum over nucleon species is implied), while the
axial current $A^\mu$ has only spatial components,
$A^i=C_A\psi^\dagger\sigma^i\psi$, where $\sigma^i$ are the Pauli matrices and
the non-relativistic $\psi$ are Pauli spinors.
If $V^i=0$ and $A^0=0$ all terms involving $\langle V^0 V^i\rangle$ or
$\langle A^0 A^i\rangle$ vanish, \ie, terms~3 and~4. Term~5 involves
components $\langle A^i V^j\rangle$ which also vanish.
Term~1 only has a $00$ component and thus corresponds to $\langle V^0
V^0\rangle$, i.e., for only one species of nucleons $R_1=C_V^2 S_\rho$ with
$S_\rho$, the usual dynamic structure function for nucleon density
fluctuations~\qref{\IwamotoP}. Term~2
only has spatial components and thus, it corresponds to
$\langle A^i A^i\rangle$, \ie\ $R_2=C_A^2S_\sigma$ with $S_\sigma$ the dynamic
structure function for spin-density fluctuations~\qref\IwamotoP.
For a mixed medium of protons and neutrons, the interpretation of $R_{1,2}$ is
more complicated because there are isospin 0 and 1 contributions to
both~\qref\SawyerII.

In the non-relativistic limit we are thus left with only two structure
functions $R_{1,2}(k)$ to consider. Turning to the long-wavelength
approximation we assume that the three-momentum transferred to the nucleons is
negligible.  This is justified by considering the perturbative series
Eq.~(\Wf).  Each contribution to $S_{\mu\nu}$ ``knows'' about the
energy-momentum transfer only through the overall energy-momentum conserving
$\delta$-function. If the nucleon mass is very large we may neglect the
momenta of the neutrinos (and other relativistic participants) in the overall
law of three-momentum conservation so that
$$
\delta^4(p)\to\delta^3(\Sigma\p_i-\Sigma\p_f)\,
  \delta(\o+\Sigma E_i-\Sigma E_f)\,.\EQ\Wj
$$
It may seem somewhat strange to drop the momentum transfer and keep the
energy transfer. Specifically,  when scattering from individual nucleons in
the non-relativistic limit, the energy transfer is smaller than the momentum
transfer:  $\omega = \k{\cdot}\p/m_N < |\k|$, where here $\p$ is the
nucleon momentum. Although this is true for single-particle scattering, and
indeed implies that $\omega = 0$ in the long-wavelength approximation, it is
not true for the higher-order terms involving the interaction of a neutrino
with a pair (or more) of interacting nucleons.
In that case, it is possible to conserve
momentum exactly in the nucleon sector while still releasing energy to or
absorbing energy from the neutrinos. Thus, in the long-wavelength limit
the perturbative response function is a $\delta$-function in $\o$ at leading
order, but at higher order the medium can transfer any amount of energy
subject to the thermal constraints of the medium and the energy available
in the incident neutrino; but to all orders it is independent of~$\k$.

Thus, in an isotropic medium there remain two structure functions
$S_V(\omega)\equiv R_1(\omega,0)$ and $S_A(\omega)\equiv R_2(\omega,0)$ which
are given in terms of field correlators as
$$
  \eqalign{S_V(\omega)&=\frac{1}{N_B}\int_{-\infty}^{+\infty} dt\,
   e^{i\omega t}\,
  \langle V^0(t) V^0(0)\rangle\,,\cr
  \alignskip
  S_A(\omega)&=\frac{1}{3N_B} \int_{-\infty}^{+\infty} dt\, e^{i\omega t}\,
  \langle \A(t)\cdot\A(0)\rangle\,,\cr}
  \EQ\SSa
$$
where
$$
  \eqalign{V^0(t)&=\int d^3\x\,
  \left[C_{V,p}\psi^\dagger_p(t,\x)\psi_p(t,\x)+
        C_{V,n}\psi^\dagger_n(t,\x)\psi_n(t,\x)\right]\,,\cr
  \alignskip
  A^i(t)&=\int d^3\x\,
  \left[C_{A,p}\psi^\dagger_p(t,\x)\sigma^i\psi_p(t,\x)+
        C_{A,n}\psi^\dagger_n(t,\x)\sigma^i\psi_n(t,\x)\right]\,,\cr}
  \EQ\SSb
$$
with the Pauli matrices $\sigma^i$.

For an isotropic distribution of neutrinos, after averaging over the
scattering or emission angles the $\cos\theta$ terms in Eq.~(\Wh) average to
zero. Therefore, we are left with $W(k_1,k_2)=\GF^2 N_B S(\omega)$ with a
single structure function
$$
  S(\o)\equiv \frac{S_V(\o)+3 S_A(\o)}{4}\,.
  \EQ\Xbb
$$
In the non-relativistic and long-wavelength limit this function is all we need
to study neutrino scattering and pair processes.



\chapter{Evaluation of $S(\omega)$}

\section{Simple Properties of $S_V$ and $S_A$}
\noindent
In general, a full evaluation of $S(\o)$ is not possible. We will
present perturbative calculations later in this section, but there are
limitations to the range of their reliability. However,
even without detailed knowledge of the structure functions we can
easily state several useful and simple properties of $S_V(\omega)$ and
$S_A(\omega)$.

To begin, the detailed-balance requirement for $W(k_1,k_2)$ translates
into
$$
  S_{V,A}(-\o)=e^{-\o/T} S_{V,A}(\o)
  \EQ\SSd
$$
so that it is enough to specify these functions for positive energy transfers
(energy absorbed by the medium).

Next, we consider the overall normalization. By integrating Eq.~(\SSa)
over $d\o/2\pi$ we extract the total strength. On the r.h.s.\
$\int e^{i\o t}\,d\o/2\pi=\delta(t)$, allowing us to trivially perform the
time integration, and find
$$
  \eqalign{\int_{-\infty}^{+\infty} \frac{d\o}{2\pi}S_V(\omega)&=\frac{1}{N_B}
  \langle V^0(0) V^0(0)\rangle\,,\cr
  \alignskip
  \int_{-\infty}^{+\infty} \frac{d\o}{2\pi} S_A(\omega)&=\frac{1}{3N_B}
  \langle \A(0)\cdot\A(0)\rangle\,.\cr}
  \EQ\SSe
$$
Of course, we could have chosen any time, not just $t=0$, as a reference
point.

If there are no spin or isospin correlations between different nucleons of the
medium we may interpret $\A$ as the single-nucleon spin operator (apart from
overall factors). In this case Eq.~(\SSe) yields the "sum rule"
$$
  \eqalign{\int_{-\infty}^{+\infty}\frac{d\o}{2\pi}S_{V}(\omega)&=
  X_pC_{V,p}^2+X_nC_{V,n}^2\equiv\SS_V\,,\cr
  \alignskip
  \int_{-\infty}^{+\infty}\frac{d\o}{2\pi}S_{A}(\omega)&=
  X_pC_{A,p}^2+X_nC_{A,n}^2\equiv\SS_A\,,\cr}
  \EQ\SSf
$$
where $X_p$ and $X_n$ are the number fractions of protons and neutrons,
respectively.\footnote{$^\dagger$}{If the medium is degenerate, but
approximated
by non-interacting particles, one must account for final
state blocking effects and replace the number fractions by
$X_i \rightarrow N_B^{-1}\int d^3\p\,2 f_i(1-f_i)/(2\pi)^3$. This result,
however, cannot be extended trivially to interacting nucleons.}

For non-interacting nucleons the time evolution of the operators is trivial,
\eg\ $\A(t)=\A(0)$, and so we find
$$
S_{V,A}(\omega)=\SS_{V,A}2\pi\delta(\omega).
\EQ\SSg
$$
This result yields directly the usual neutrino scattering rate in a medium of
non-relativistic, non-degenerate, quasi-free nucleons:
$\Gamma=\GF^2 N_B \o_1^2 (\SS_V+3\SS_a)/4\pi$,
where $\o_1$ is the energy of the incident neutrino.

It is expected that the
normalization and
shape of the response
functions will be altered in an interacting medium, however this primarily
affects the axial response $S_A$. For the vector current Eq.~(\SSg) should
be a good approximation even in the presence of nucleon-nucleon interactions
because the quantity $V^0$ (the zeroth component of the four velocity) does
not fluctuate; \ie\ it is identically~1 in the non-relativistic
limit where the wavelength of the momentum exchange is large compared to
the smearing of the nuclear charge density due to interactions in the
medium. The axial charge, on the other hand, is a spatial 3-vector quantity.
Although a nucleon's location is not smeared out, nucleon interactions in the
medium cause the direction of the nuclear spin to fluctuate, altering the
effective charge seen by a leptonic probe.
We expect that the axial response is weakened for soft
energy transfers, $\o < \gamint$, where by $\gamint$ we mean some
typical rate for the nuclear spin to fluctuate due to
collisions.\footnote{$^\dagger$}{Isospin
fluctuations do not lead to an equivalent damping of the vector response,
due to isospin conservation in the strong interactions. In order to have the
equivalent effect a proton must turn into a neutron, but this can only
occur if a neutron turns into a proton in the same interaction. In the
long wavelength approximation, the net weak vector charge probed by the
neutrinos will remain unchanged.}

The counterpoint to these arguments is that there is no vector response for
non-zero energy transfers, but in general the axial response gets spread
out to higher $\o$ with increasing interactions. Specifically, explicit
calculations (see \S~III.b below) show that only the axial-vector current
contributes to bremsstrahlung in the non-relativistic limit.

A related issue concerns the appropriate values to use for the axial and
vector charges $C_A$ and $C_V$. In a dilute medium, one should use the
vacuum values, but in a nuclear medium the axial weak coupling constants
may be altered. The vector charges remain unchanged. Specific values are
discussed in Appendix~A.

Thus, within the non-relativistic and long-wavelength approximation
the vector current is unaffected by multiple collisions but
does not contribute to bremsstrahlung, and the reverse applies
to the axial-vector current. It remains to determine the
axial-vector structure function $S_A(\omega)$ in a fully interacting
medium.

Finally, it should be clear that random spin fluctuations of the sort
which happen in a thermal medium may lessen the total scattering
interaction rate, but we would be surprised if any significant enhancement
could occur through thermal interactions. Thus, the normalization of $S$
in Eq.~(\SSf) should constitute a rough upper bound. This does not allow,
however, for collective effects such as spin-waves, or the influence of a
pion condensate, which could in principle enhance the interaction rates of
neutrinos with the medium.

\section{Perturbative Result for $S(\omega)$}
\noindent
We proceed with a perturbative calculation of $S_A(\omega)$. After the
$\delta$-function contribution from free nucleons, the next
lowest-order process that can contribute is nucleon-nucleon scattering.
We consider non-degenerate nucleons and model the interaction by one-pion
exchange.  Neglecting $m_\pi$ in the pion propagator we extract from Friman
and Maxwell's matrix element~\qref\FrimanM\ and from Turner's emission rate
for the bremsstrahlung process $NN\to NN\overline\nu\nu$~\qref\Turner\ the
result
$$
  S_\NN(\omega)=C_A^2\,\frac{\Gamma_A}{\omega^2}\,s(\omega/T)\,.
  \EQ\Yc
$$
where the NC axial coupling $C_A$ is, for simplicity,
taken to be the same for neutrons and protons, and
$$
  \Gamma_A \equiv f(X_n)\,48\,\pi^{1/2}\,
  \frac{\alpha_\pi^2\,N_B\,T^{1/2}}{m_N^{5/2}} \,.
  \EQ\Ycc
$$
Further, $f(X_n)=1+4X_n(1-X_n)$, and
$\alpha_\pi\equiv(f 2m_N/m_\pi)^2/4\pi\approx 15$, where $f\approx1$ is the
pion-nucleon coupling. The function
$$
  s(x)\equiv
  \int_0^\infty dy\,\left(y^2+ x y\right)^{1/2} e^{-y}
  \approx\left(1+\frac{\pi}{4}\,x\right)^{1/2}
  \EQ\Yccc
$$
with $x=\o/T$ is the remnant of the nuclear matrix element after doing the
phase space integration for the nucleons. The analytic approximation has a
maximum error of less than $2.2\,\%$.

This result cannot be the complete answer, however, as it diverges at
$\omega=0$.
The perturbative calculation is based on the assumption that a given
bremsstrahlung event can be viewed as a single collision where in-states
travel unperturbed from the infinite past and are scattered into out-states
which remain unperturbed into the infinite future.
In a medium these assumptions are
violated as each nucleon participant in a given reaction has emerged from a
previous interaction and will interact again in the future. The nucleon field
can freely evolve only for an approximate duration $\tau=\gamint^{-1}$ where
$\gamint$ is a typical nucleon-nucleon scattering rate. Hence, we cannot
expect such a calculation to yield meaningful results for energy scales
$\vert\omega\vert\lapprox \tau^{-1}=\gamint$.
This estimate reveals the approximate range of $\o$ where a naive
perturbative approach can be trusted.

In a non-degenerate medium in thermal equilibrium, the scale of a typical
energy
transfer in any collision or radiation event is given by the temperature $T$.
Hence, as long as $T/\gamint \gg 1$ one may expect that a naive application of
pertubative methods is justified and we call such a medium ``dilute'' for the
purposes of our discussion. Conversely, the
criteria $T/\gamint \ll 1$ quantifies our notion of a ``dense''
medium (the limit of large collision rates) where the
naive use of perturbative results is problematic because it is not justified in
any obvious way. The hot material of a young SN core appears to be
a medium which is approximately non-degenerate and yet ``dense'' in
the spirit of this definition.

Energy conservation in the collisions between neutrinos and nucleons will be
satisfied only with a precision $\Delta E\approx\gamint$. This observation
suggests that the nucleons may be described as resonances with an approximate
width $\gamint$, rather than as states with a fixed energy~\qref\Weldon. In a
previous paper we have argued~\qref\RaffeltSS\ that in this case one should
replace the $1/\omega^2$ soft behavior of $S_A(\omega)$ by a Lorentzian
$1/(\omega^2+\gamint^2/4)$. If we use this form for the soft part of
$S_A(\omega)$, and identify $\gamint = \frac{4}{3}\Gamma_A$, we may write
$$
  S'_\NN(\omega)= C_A^2 \frac{\Gamma_A}{\omega^2+
 \frac{4}{9}\Gamma_A^2}\,s(\omega/T)\,.
  \EQ\Yd
$$
This form has two attractive features in the limit of a dilute, but
interacting, medium. For small values of $\Gamma_A$
it approaches the appropriate $\delta$-function for $S_A$ while for
$\omega\gg\Gamma_A$ the correct bremsstrahlung emissivities obtain.

\section{Large Collision Rates}
\noindent
We have argued that in the limit of small collision rates, defined by the
condition $\Gamma_A\lapprox T$, the low-$\omega$ behavior can be reasonably
approximated by a Lorentzian shape. In a supernova core, however, we are
confronted with the opposite case of large collision rates. Numerically,
Eq.~(\Ycc) becomes
$$
  \frac{\Gamma_A}{T} \approx 415 \, \frac{\rho}{\rho_0}\,
  \left(\frac{800\,\MeV}{m_N}\right)^{5/2}
  \left(\frac{50\,\MeV}{T}\right)^{1/2}
  \left(\frac{f(X_n)}{1.84}\right)\,,
  \EQ\Ye
$$
where $\rho_0=3{\times}10^{14}\,\rm g/cm^3$ is the nuclear density, and we
have scaled the factor $f(X_n)$ to its value at $X_n = 0.3$.
The numerical value of $\Gamma_A/T$ is approximately a factor 10 larger
than the value we estimated for $\gamint / T$ in a previous
paper~\qref\RaffeltS.
That estimate was based on low energy $p$-$p$ and $p$-$D$ scattering data,
which may or may not be appropriate as a model for the source of spin
fluctuations in a medium.
It is the enormous magnitude of $\Gamma_A/T$ that makes the interpretation of
the perturbative results for $S_A$ very problematic.

If we used the form Eq.~(\Yd) then both neutrino scattering and
pair processes would be substantially suppressed. This will be shown
explicitly in \S~IV. For now, it suffices to note that in a thermal
environment, typical values of $\o$ are a few $T$, and if $\Gamma_A \gg T$
then the denominator in Eq.~(\Yd) becomes large for reasonable $\o$.
Although there may be scattering strength at large $\o$ that strength will
not be sampled by thermal energy transfers.

It is plausible, however, that in a dense medium $S_A(\o)$ is
narrower than indicated by a naive extrapolation of the dilute-medium result.
The width of the spin-spin correlation function for a single nucleon
is determined essentially by scattering off the spatial spin fluctuations in
the medium. But if those spins are all fluctuating rapidly in time, then
{\it their} spatial fluctuations will be damped by the same mechanism.
To get a consistent picture, one must self-consistently include the effects of
interactions in an evaluation of the interaction rates.

Even though it appears impossible to extract reliable results for $S$ from a
simple-minded perturbative calculation, we believe that the overall shape
of the structure function will be a broad distribution, and that the high-$\o$
wings will be represented by the bremsstrahlung result. However, it is in
no way evident what one should use for the soft part of the axial response
function. The Lorentzian shape espoused in Eq.~(\Yd) is possible, but it
is also possible that some radically different shape is correct.
Indeed, in the next section we will show that extreme changes to the
shape are required if pair processes are to dominate over scattering
processes.



\chapter{Scattering vs.\ Pair Processes}

\noindent There are many ways that one could choose to compare the strength of
the scattering and pair processes. For example, we could concentrate on the
contribution to various transport coefficients or calculate the Rosseland
mean opacity.  To illustrate our arguments we will use the rate at which
particles are absorbed by either the pair or scattering terms, averaged over
the neutrino phase space,
$$
  \eqalign{
  \gamscat &= \frac{1}{N_\nu} \int \frac{d^3\k_1}{(2\pi)^3}
  \int\frac{d^3\k_2}{(2\pi)^3} \,W(k_1,k_2)\, f(\o_1)\,[1-f(\o_2)]\,,\cr
  \alignskip
  \gampair &= \frac{1}{N_\nu} \int\frac{d^3\k_1}{(2\pi)^3}
  \int\frac{d^3\k_2}{(2\pi)^3} \,W(k_1,-k_2)\, f(\o_1) \,
  \overline f(\o_2)\,.\cr}
  \EQ\Xa
$$
These quantities have the advantage of being simple to calculate and
interpret, without being too specific to the details in some corner of phase
space.  We have normalized by the number density of neutrinos
$N_\nu = \int f(\o_1) d^3\k_1/(2\pi)^3$.  The occupation numbers will be taken
to be Fermi-Dirac distributions without a neutrino chemical potential.

In \S~II we argued that for the conditions relevant for a SN core $W(k_1,k_2)$
depends only on one function $S$ of the energy transfer $\o$ according to
Eq.~(\Xbb). Therefore,
$$
  \gamscat = \norm
  \int_0^{\infty} d\omega_1\, \int_0^{\infty} d\omega_2\,
  \omega_1^2\, \omega_2^2\, f(\o_1)\, [1-f(\o_2)]\, S(\omega)\,,
  \EQ\Xd
$$
or transforming to an integral over $\otil=\o_2-\o_1=-\o$
$$
  \gamscat = \norm
  \int_{-\o_1}^{\infty} d\otil\, \int_0^{\infty} d\omega_1\,
  \omega_1^2\,(\otil + \o_1)^2\, f(\o_1)\,[1-f(\otil{+}\o_1)]\,S(-\otil)\,.
  \EQ\Xdd
$$
By means of the detailed-balance requirement $S(-\o) = e^{-\o/T}\,S(\o)$, and
relabeling the integration variable $\otil\to\o$, we may write this as an
integral over positive $\o$ only,
$$
\gamscat = \int_0^\infty d\omega\, \fscat(\omega) S(\omega)  \,.
  \EQ\Xf
$$
We have arranged things so that we will need $S(\o)$ only for positive energy
transfers (energy absorbed by the medium). The detailed-balance Boltzmann
factor $e^{\o/T}$, relevant for negative energy transfers (energy given to the
neutrinos) has been included in the definition of the phase-space factor
$F(\o)$ which is specific to a given process and chosen type of thermal
average, \eg\ by number, by energy, etc.

$\fscat$ is then expressed as a sum of two terms because we split the $\o$
integration to ensure that $\o$ is always positive,
$$
\eqalign{\fscat (\omega) = \norm
 \Biggl\{&\int_{\omega}^{\infty}\!\!\! d\omega_1\,
   \o_1^2 (\omega_1-\omega)^2\, f(\o_1)\,[1- f(\o_1{-}\o)] \cr
 \alignskip
  & + \int_{0}^{\infty}\!\!\!  d\omega_1\, \o_1^2 (\omega_1+\omega)^2\,
   f(\o_1) [1-f(\o_1{+}\o)] e^{-\o/T}\,\Biggr\}\,.\cr}
\EQ\Xg
$$
Note that in the first term $\o = \o_1 - \o_2$, while in the second $\o = \o_2
- \o_1$.  Also, there is a detailed-balance factor $e^{-\o/T}$ in the second
integral relevant for the case when the medium gives energy to the neutrinos
rather than absorbing it. Finally, in the case $S(\o) \sim \delta(\o)$
only one of the terms in Eq.~(\Xg) should be included, to avoid double
counting.

Similar manipulations lead to
$\gampair = \int_0^\infty d\omega\, \fpair(\omega) S(\omega) $
with
$$
  \fpair(\omega) = \norm
  \int_{0}^{\omega}\!\!\! d\omega_1\,
  \o_1^2 (\o-\o_1)^2\,  f(\o_1)\,\overline f(\o{-}\o_1)\,.
  \EQ\Xgg
$$
Because we consider thermal neutrino distributions,
$F_\pair$ equally applies to the phase-space averaged pair emittance.

As a further simplification, for non-degenerate neutrinos we may approximate
the Fermi-Dirac by Boltzmann distributions, and ignore the Pauli-blocking
factor in Eq.~(\Xa). In $\fscat$ and $\fpair$ this means that
$f(\o_1) \to e^{-\o_1/T}$ and $\overline f(\o_2) \to e^{-\o_2/T}$ while
$1-f(\o_2)\to 1$.  Then we find analytically,
$$
\eqalign{\fscat(x) &\approx \frac{\GF^2 N_B T^2}{2\pi^2}\,
  \left(24+12x+2x^2\right)\,e^{-x}\,,\cr
  \alignskip
  \fpair(x) &\approx \frac{\GF^2 N_B T^2}{2\pi^2}\, \frac{x^5}{60}\,e^{-
  x}\,,\cr}
  \EQ\Xh$$
where $x\equiv\omega/T$.
These approximations are shown in Fig.~\fF; they deviate from
Eqs.~(\Xg) and (\Xgg) by less than 10\%, except for $\fpair$ at
values of $x$ less than 5, where Eq.~(\Xh) overestimates $\fpair$ by up to
a factor of 3. Since $\fpair \ll \fscat$ for these $x$, we can use Eq.~(\Xh)
without reservation.

{}From Fig.~\fF\ it is clear that scattering is more sensitive to the structure
function near $\omega=0$ while pair absorption is more sensitive to $S$ at
higher $\omega$. Put another way, for pair processes to dominate, $S(\omega)$
would need to have much more power at large $\omega$ than near $\omega=0$. For
example, if $S(\omega)$ would vary like a power law $\omega^n$, $\gampair$
would exceed $\gamscat$ for $n>3$, but would be subdominant for smaller $n$.

For a non-interacting medium one can use the $\delta$-function result
 Eq.~(\SSg).
This limit allows for no pair processes, and
scattering dominates. The scattering rates due to this choice of $S(\o)$
are shown as solid horizontal lines in Fig.~\fGamma. Including
interactions broadens the axial response. For a dilute
interacting medium, the Lorentzian shape, Eq.~(\Yd) seems a reasonable
approximation. For large $\o$, the axial response then behaves as
$\o^{-2}$, a result which is much softer than the $n>3$ power law required
for pair processes to dominate. Again, there is no scope for pair
processes to dominate.

In the limit of large collision rates there is more room for speculation.
In order to avoid the conclusion that $\Gamma_\scat>\Gamma_\pair$
we would have to assume that $S(\o)$ is not monotonically
decreasing, but rather is ``hollowed out'' near $\o=0$. A radical example
would be to take Eq.~(\Yc) for $S(\o)$ above some cutoff $\omin$, and to set
$S(\o)=0$ for $\o<\omin$.\footnote{$^\dagger$}{
For the sake of this extreme example we ignore the normalization of
Eq.~(\SSf). If that relation is indeed an upper bound then the implied
cutoff is $\omin \gapprox 10^5$. Such a large value suggests that some
other form of regulation must apply, \eg\ as in the Lorentzian form
$S'_\NN$.}
Under this assumption, the rates
$\Gamma_\scat$ and $\Gamma_\pair$ are shown as the dotted curves in
Fig.~\fGamma. As can be seen, $S_\NN$ becomes more important for pair processes
than for scattering for $\omin/T\gapprox 5$. However, even if such a radical
transformation of the axial response occurs, there will still be the
vector contribution, $S(\o) = 2 \pi \SS_V \delta(\o)$, to scattering.
{}From Fig.~\fGamma, for $\rho = \rho_0$ and $\omin/T \gapprox 6$,
vector current scattering dominates over pair processes.
Thus, even with the radical assumption of a total cutoff of the low-$\o$
response there is almost no range where pair processes dominate.

Not only must the axial response be hollowed out, but in order for the
pair processes to dominate, the axial strength at high $\o$ must remain strong.
At the same time, we note that for $\omin/T \lapprox 4$, the scattering
response
is enhanced over the single nucleon result, a consequence of the
$\o^{-2}$ behavior of $S_\NN$. If we use the Lorentzian form, the
$\o^{-2}$ behavior is regulated, but the strength of the axial scattering
is severely suppressed. For example, using Eq.~(\Yd),
for $\o > \omin$ instead of the naive bremsstrahlung result, Eq.~(\Yc),
yields the two dashed curves in Fig.~\fGamma. Again, pair processes exceed
scattering for a cutoff $\omin/T \approx 5$, however now both processes are
two orders of magnitude smaller than the vector contribution.

The results shown in Fig.~\fGamma\ are for $\rho = \rho_0$, and
$\Gamma_A \approx 415$. As the density is raised even higher, the
single particle contributions to the rates increase linearly with density,
while the naive bremsstrahlung rates increase as $\rho^2$. So it would
appear that eventually pair processes could overtake the contribution from
$S_V$. However, if we look at the contribution from $S'_\NN$ instead of
$S_\NN$ we find that the contribution to $\Gamma$ is independent of
$\rho$, and the comparison favors scattering even more strongly.
Given the wide range of possibilities for $S$, it is not clear whether
or not increasing the density can eventually allow pair processes to win out.

We do not think that an extreme behavior such as a sharp cutoff to $S(\o)$
is realistic, so we view these exercises as examples of how radically
different the high-density behavior of $S$ would have to be in order that
pair processes might dominate.

We conclude that, even though we cannot calculate the exact form of the
axial response of the medium, a combination of phase space arguments and a
recognition of the stability of the vector response, seems to preclude pair
processes dominating over scattering processes in the evaluation of
neutrino transport properties. In order to avoid this conclusion the axial
response function for a strongly interacting medium would have to be both
suppressed at low and enhanced at high energy transfers in comparison to
the bremsstrahlung result, $S_\NN$.

Another effect of increasing density is degeneracy of the
nucleons. As the nucleons become degenerate the amount of nucleon phase
space available for small energy transfers decreases, effectively creating
a harder response function.
{}From Friman and Maxwell's~\qref\FrimanM\ bremsstrahlung rate we
can easily extract $S_\NN$. The result includes various prefactors,
but for our purposes the interesting part is the dependence
on the energy transfer,
$$
s_{\rm deg}(x) = \frac{x^3+4\pi^2 x}{x^2 + \gamma^2/4}
  \,\,\frac{e^x}{e^x - 1} \,,\EQ\Ycc
$$
where $x = \o/T$ and $\gamma=\gamint/T$ is our phenomenological cutoff,
again necessary to regulate an $x^{-2}$ divergence at small $\o$.  Now,
however, the large $x$ behavior is much ``stiffer'', $s_{\rm deg}\propto x$.
This is still not hard enough to provide for the dominance of pair processes,
but it is getting closer. In the extreme case where $\gamint \gg T$,
the effective power law of the response function is $n = 3$.  This last value
is enough to make pair processes competitive; however, we don't think this
extreme case really applies to the SN core since the temperatures are high
enough that the nucleons are only mildly degenerate.

In summary, we repeat that it is in no way evident how one ought to
extrapolate the dilute-medium approximation for $S_A(\o)$ into the high-density
regime. However, unless the overall shape of $S$ is radically different from
its limiting behavior,
the pair-processes will always be less important than scattering. Further, it
appears that both the pair-rates and the axial-vector scattering rates will be
suppressed relative to their ``naive'' values.

Finally, one may wonder if these conclusions are not in conflict with
the reasonable agreement between the SN 1987A neutrino observations and the
expected signal duration. However, as the vector-current appears to remain
unsuppressed, the scattering rates would be reduced only to about $1/5$ of
their naive values, even if the axial-vector contributions were entirely
suppressed. It is an interesting question if, given the freedom to adjust
other parameters, such a reduction in neutrino opacities can be
excluded on the basis of the detected events.

\chapter{Non-Standard Physics}

\noindent
We now turn to exploring how our conclusions affect various arguments related
to certain non-standard aspects of SN physics. First, we consider the emission
of hypothetical particles weakly coupled to the NC of the
medium. This process could provide an anomalous cooling mechanism for the
core, which in turn could result in a diminished neutrino signal of SN~1987A,
thus allowing one to constrain the properties of the particles in
question~\qref\Raffelt. In fact, trying to understand the consistency of such
arguments was the original motivation for this work. Second, we look at
the possible impact of a pion condensate for both neutrino transport and
hypothetical particle emission.  All these issues are connected through the
same medium structure functions that control neutrino transport. It is
therefore possible, and necessary, to treat the novel effects and neutrino
transport in a consistent fashion.
\bigskip

\section{Processes With ``Flipped'' Neutrinos}
\noindent
As a first case we consider the possibility that neutrinos have a small Dirac
mass $m_\nu$, which allows for the production of ``wrong-helicity'' states in
the deep interior of a SN core~\qref\RaffeltS.  If $m_\nu$ is small enough
these states can escape freely and thus provide an anomalous sink for the heat
in the core. If $m_\nu$ is not too small then that heat sink would have had
observable effects on the SN 1987A neutrino signature, thus constraining
$m_\nu$. Several estimates of these bounds have been made, often with
different treatments of the processes which contribute to the
emission~\qref{\Turner,\BurrowsGT,\Wilson,%
\LamN,\GaemersGL,\RaffeltS,\GandhiB}.  Specifically, under certain
assumptions, the strength of the pair emission processes
$X\to X'\overline\nu_L\nu_R$ has been shown to
exceed the rate for spin-flip scattering from single nucleons
$\nu_L N\to N'\nu_R$ \qref{\Turner,\BurrowsGT,\Wilson}.
In analogy to our remarks about the importance of pair processes in the
transport of ordinary l.h.~neutrinos; we argue that unless the full response is
very ``hard'' spin-flip scattering still dominates over pair emission.

We take $m_\nu$ to be very small compared with typical neutrino energies.
Therefore, it is not necessary to distinguish carefully between helicity
and chirality, and so we shall always refer to wrong-helicity neutrinos or
anti-neutrinos as right-handed and to the correct-helicity states as
left-handed.  Moreover, if $m_\nu$ is sufficiently small, r.h.~states
will not be trapped so that their occupation numbers are sub-thermal.  The
energy-loss rate per unit volume due to the emission of r.h.~neutrinos is then
$Q_{\nu_R}=Q_\scat+Q_\pair$ with
$$\eqalign{Q_\scat&=
  \int\frac{d^3\k_L}{(2\pi)^3}\,\frac{d^3\k_R}{(2\pi)^3}\,
  \widetilde W_{k_L,k_R}\,f_{\k_L}\,\o_R \,,\cr
  \alignskip
  Q_\pair&=
  \int\frac{d^3\k_L}{(2\pi)^3}\,\frac{d^3\k_R}{(2\pi)^3}\,
  \widetilde W_{-k_L,k_R}\,(1-\overline f_{\k_L})\,\o_R \,,\cr}
  \EQ\Qa$$
where $\widetilde W_{k_L,k_R}$ is the transition probability for the
scattering process $\nu_L(k_L)+X\to X'+\nu_R(k_R)$ while
$\widetilde W_{-k_L,k_R}$ refers to $ X\to X'+\overline\nu_L(k_L)+\nu_R(k_R)$.
For the energy loss due to $\overline\nu_R$ one obtains an analogous
expression; for non-degenerate $\nu_L$ we have $Q_{\overline\nu_R}=Q_{\nu_R}$.

In order to determine $\widetilde W_{k_L,k_R}$ we note that it can be written
in the form Eq.~(\Wc) where the structure function $S_{\mu\nu}$ remains
unchanged.  For interactions of neutrinos with specified helicities, Gaemers,
Gandhi, and Lattimer~\qref\GaemersGL\ showed that the expression for
$N^{\mu\nu}$ remains of the form Eq.~(\Wd) if one substitutes
$k_i\to\half(k_i\pm m_\nu s_i)$, $i=1$ or~2, where the plus-sign refers to
$\nu$ and the minus-sign to $\nubar$, and $s$ is the covariant spin-vector.
For relativistic neutrinos we may consider a non-covariant lowest-order
expansion in terms of $m_\nu$. Then $k_i$ remains unchanged for l.h.\ states
while
$$k_i=(\o_i,\k_i)\to\tilde k_i=(m_\nu/2\o_i)^2\,(\o_i,-\k_i) \EQ\Qb$$
for r.h.\ ones. After this substitution has been performed, all
further effects of a non-zero $m_\nu$ are of higher order so that
one may neglect $m_\nu$ everywhere except in the global spin-flip
factor.\footnote{$^\dagger$}{The dispersion relation of neutrinos in a SN
differs markedly from the vacuum form; in the core the ``effective $m_\nu$''
is several $10\,\keV$.  However, $m_\nu$ in equation~(\Qb) is the vacuum mass
which couples l.h.\ and r.h.\ states and thus leads to spin flip while the
medium-induced ``mass'' only affects the dispersion relation of l.h.\ states.
This view is supported by a detailed study of
Pantaleone~\qref\Pantaleone. Of
course, for non-relativistic neutrinos the situation is more complicated
because an approximate identification of helicity with chirality is not
possible.}
Therefore, the neutrino tensor $\widetilde N^{\mu\nu}$ is found from Eq.~(\Wd)
by multiplication with $(m_\nu/2\omega_R)^2$, inserting $k_1=k_L$, and
$k_2=(\omega_R,-\k_R)$.

Next, we explicitly contract $\widetilde N^{\mu\nu}$ with $S_{\mu\nu}$ in the
form Eq.~(\Wg) in order to derive the spin-flip equivalent of Eq.~(\Wh),
$$
\eqalign{
  \widetilde W(k_L,k_R)=& \frac{\GF^2 N_B}{4}\,
  \left(\frac{m_\nu}{2\o_R}\right)^2\,
  \biggl[(1-\cos\theta)\,R_1+ (3+\cos\theta)\,R_2 \cr
  \alignskip
 & +4\o_R^2(1-\cos\theta)\,R_3
  -4\o_R(1-\cos\theta)\,R_4 +2(\o_R-\o_L)(1+\cos\theta)\,R_5 \biggr]
  \,.\cr}\EQ\Qc
$$
Following Gaemers, Gandhi, and Lattimer~\qref\GaemersGL\ we emphasize that
this expression differs in more than the factor $(m_\nu/2\o_R)^2$ from the
non-flip rate Eq.~(\Wh). This difference is related to the changed angular
momentum budget of reactions with spin-flipped neutrinos.

In the non-relativistic limit, however, the contributions of $R_{3,4,5}$ can
be neglected as discussed in \S~II.c above. Moreover, in an isotropic medium
the $\cos\theta$ terms average to zero. Then, the spin-flip factor is the only
modification necessary to deal with r.h.~neutrinos, and we may write
$\widetilde W(k_L,k_R)=\GF^2 N_B (m_\nu/2\o_R)^2\,S(\omega)$, which involves
the same structure function as for the non-flip case. Then, in analogy to
Eq.~(\Xf), the energy-loss rates can be written as
$$
  Q_i = \int_0^\infty d\o \widetilde F_i(\o)\,S(\o)\,,\EQ\Qd
$$
where, $i$ stands for ``pair'' or ``scat''.  If for non-degenerate neutrinos
we use a Boltzmann distribution and neglect Pauli-blocking in Eq.~(\Qa), the
phase-space functions are
$$
  \eqalign{\widetilde F_{\rm scat}(x)&=\frac{\sigma_\flip N_B T^4}{4\pi^3}\,
  \left(x^2+6x+12\right)\,e^{-x}\,,\cr
  \alignskip
  \widetilde F_{\rm pair}(x)&=\frac{\sigma_\flip N_B T^4}{4\pi^3}\,
  \frac{x^4}{12}\,e^{-x}\,,\cr}
  \EQ\Qe
$$
where $\sigma_\flip\equiv\GF^2 m_\nu^2/4\pi$ and $x=\o/T$.

A comparison of Eqs.~(\Qe) and (\Xh) shows that the r.h.~neutrino
emission bears a resemblance to l.h.~neutrino absorption: in order for
pair processes to dominate, the response function must be very hard.
In \S~III.c we characterized the ``hardness'' of $S$ by a power law $\o^n$,
and noted that the ``critical exponent'' where pair processes were as
important as scattering was $n=3$. Similarly, for
r.h.~neutrino emission, the critical exponent is $n \approx  2.57$.

Since the critical spectrum for spin-flip pair emission is slightly softer
than for l.h.~pair absorption, it is slightly easier for
pair-processes to dominate, than in the case of l.h.~neutrino transport.
In Fig.~\fQ\ we show the r.h.~emission results analagous to the
l.h.~scattering results shown in Fig.~\fGamma. The emission rates are scaled
to $Q_0 = \sigma_\flip N_B T^4 /(4\pi^3)$. The results are similar to
those presented for the transport of l.h.~neutrinos. The cross-over to
pair processes dominating the axial emission occurs at
$\omin/T \approx 4$, but a combination of axial plus vector scattering
almost always dominates even when no attempt (other than the cutoff at
$\omin$) is made to regulate the soft $\o$ response. When the Lorentzian
form for $S_A$ is used, vector scattering dominates at nuclear density.
Therefore, by any indication that we can obtain from extrapolating
perturbative calculations into the high-density regime, the scattering
processes are dominant for spin-flip emission of r.h.~states, as well as
for the transport of l.h.~neutrinos.

Even though we can be reasonably sure that scattering processes dominate
the emission rate of r.h.~neutrinos there is still uncertainty about the
total emission rate due to the uncertainty about the axial contribution.
If the axial contribution is much suppressed then $Q_{\scat, V}$ should
give a good lower bound to the emission rate, whereas if the axial part is
enhanced the emission rate may be much larger than $Q_{\scat, V}$. We feel
that the latter possibility is remote but cannot be ruled out on the basis
of perturbative calculations alone.

Even though perturbative calculations do not seem to allow for a reliable
calculation of the absolute magnitude of neutrino scattering rates, the
NC transport of ordinary l.h.~neutrinos and the spin-flip
emission of r.h.~Dirac-mass neutrinos are linked to each other
through a single response function. In fact, the functions $F_\scat(x)$ and
$\widetilde F_\scat(x)$ in Eqs.~(\Xh) and~(\Qe) even have identical
shapes.\footnote{$^\dagger$}{If one of $\nu_\mu$ or $\nu_\tau$ had a large
enough mass so that spin-flip processes would be important at all, and if it
had a not-too-small mixing angle with $\nu_e$, a degenerate sea of this flavor
would be populated~\qref{\Turner,\Maalampi}. For such a degenerate flavor the
phase-space functions $F_\mu(\omega)$ and $\widetilde F_\mu(\o)$ for ordinary
and spin-flip scattering would be different because the former would involve
neutrino Pauli-blocking factors for the final-state $\nu_L$. The two
processes would then involve different integrals over the common function
$S(\o)$.}
Because modifications of the transport coefficients and of the spin-flip rates
would both affect the observable neutrino light curve, a consistent attempt to
constrain Dirac masses would then depend on letting the NC transport
coefficients and the spin-flip emission rate ``float'' together.

\section{Axions}
\noindent The emission of axions is another interesting possibility for an
anomalous energy sink in the inner core of a SN. Again, the expected impact on
the SN 1987A neutrino signature was used to constrain the coupling strength
and then indirectly the mass of these pseudo-scalar
bosons~\qref\AxReview. They would couple
to the medium according to
$$
  H_{\rm int}=\frac{1}{f_a}\, B^\mu_a \partial_\mu \phi\,,
  \EQ\YYYa
$$
where $\phi$ is the axion field and $f_a$ is the axion decay constant
which has dimensions of energy. The medium current is presumably dominated by
protons and neutrons according to
$$
  B_a^\mu=\sum_{j=n,p} C_j\,
  \overline\psi_j\gamma^\mu\gamma_5\psi_j\,,
  \EQ\YYYb
$$
where the $C_{n,p}$ are model-dependent dimensionless coupling constants of
order unity.
The structure of this interaction is of the current-current type so that we
may apply an analysis similar to that for neutrino processes, except that
we replace $N^{\mu\nu}$ by its axionic equivalent
$\Phi^{\mu\nu}\equiv\ka^\mu \ka^\nu/2\o_a$, where $\ka$ is the four momentum
of the axion and $\o_a$ its energy. Note that by our definition of the
momentum transfer $k = -\ka$.

If axions are weakly enough coupled they are not trapped in a SN core, so that
their occupation numbers are sub-thermal. In this case the energy-loss rate
per unit volume is
$$
  Q_a=\frac{1}{f_a^2}\int \frac{d^3\k}{(2\pi)^3}\,\o_a\,
  S_a^{\mu\nu}\Phi_{\mu\nu}\,,
  \EQ\YYYc
$$
The response function $S_a^{\mu\nu}$ is
as defined in Eq.~(\We) with the above current $B_a^\mu$.  In the
non-relativistic and long-wavelength limit we have
$$
  Q_a=\frac{N_B}{4 \pi^2\,f_a^2}\,
  \int_0^\infty d\o\,\o^4\,e^{-\o/T}\,S_a(\o)\,,
  \EQ\YYYd
$$
where $S_a = R_{2,a}$.
Apart from normalization the phase-space function for axion emission,
$\o^4 e^{-\o/T}$, is identical to that for spin-flip pair emission.
Note that $R_{1,a}$ vanishes because the current
$B_a^\mu$ is purely axial. Since the axion couplings to the nucleons are
generally different from the axial current couplings of the nucleons in the
weak interactions, the isospin structure of $R_{2,a}$ will differ in
detail from that of $R_2$ for neutrinos. However, the general
considerations behind a calculation of $R_2$ are entirely analagous in the
two cases. Specifically, perturbative calculations done using the one-pion
exchange approximation show different functional dependences on the
proton and neutron fractions, \ie\ the analogs to $f(X_n)$ in Eq.~(\Ycc),
but the same sort of soft divergence in the energy transfer $\o^{-2}$.
Therefore, we may use the neutrino response function $S_A(\o)$ as
a template to discuss the properties of $S_a(\o)$, without worrying too
much about the numerical accuracy of the prefactor.

{}From our previous discussions several points should be evident. First,
naive axion bremsstrahlung calculations rely on an $S_a$ similar in form
to the $S_{NN}$ given in Eq.~(\Yc). Although there is no divergence in the
emission rate, that is only because the phase space factor has sufficient
powers of $\o$ to regulate the divergence. However, since that divergence
does show up in neutrino transport, it must be regulated in some fashion;
\eg, the modification to the Lorentzian form in Eq.~(\Yd), or the
sharp cutoff at $\omin$ used in Figs.~\fGamma\ and \fQ. It seems likely that
whatever mechanism is chosen to regulate the divergence at small $\o$,
that regulation will result in a suppresion of the axion emission rates
compared to the naive bremsstrahlung calculations,
perhaps by many orders of magnitude. For
example, the effect of the Lorentzian regulation is between 3 and 4 orders
of magnitude at nuclear density, as can be seen by comparing the curves
labeled $Q_{\pair, NN}$ and $Q'_{\pair, NN}$ in Fig.~\fQ.
Conversely, if one were to insist that the naive calculation of the axion
emission rates were correct, one would have to conclude that the neutrino
scattering rates are much enhanced,
in violation of our expectation that thermal spin fluctuations can
only decrease the net interaction rate.
To probe the effects axion emission
would have on the neutrino cooling curves of supernova cores, one must
treat axion emission and neutrino transport using the same assumptions
about the axial response functions of the medium.

\section{Pion-Induced Processes}
\noindent
It is also interesting to consider non-standard effects in the nuclear medium
that could grossly affect SN cooling. In particular, the conditions are close
to forming a pion condensate, and in some equations of state the number of
$\pi^-$ can be much in excess of a thermal distribution~\qref{\Glendenning}.
A pion condensate can have an important impact on the late-time cooling of
neutron stars~\qref{\Pioncondensate}, and it was recently speculated that
under certain assumptions it could have a substantial effect on neutrino
processes in a newborn neutron star also~\qref{\Turner,\Wilson}.

We calculate the one-pion contribution to the structure function on the basis
of $\pi^- p$ in the hadronic initial state, and $n$ in the final state, while
we neglect $\pi^0$ and $\pi^+$ as we are mostly interested in the case of a
$\pi^-$ condensate. For the pion-nucleon interaction we use the pseudoscalar
form
$$
H_{\pi np} = \sqrt{2}\,\frac{2 f m_N}{m_\pi}\,
  \overline\psi_n \gamma_5 \pi^- \psi_p \,.
  \EQ\Za
$$
We set $m_n = m_p = m_N$, and keep $\alpha_\pi=(2 f m_N/m_\pi)^2/4\pi$
constant. Moreover, we calculate in the non-relativistic and long-wavelength
limit, leading to $\delta^3(\k_p - \k_n)$ for momentum conservation. Thus,
$E_n=E_p$ and we are left with $\delta(\opi+\o)$ for energy conservation.
With these simplifications we find for the one $\pi^-$ contribution
to~Eq.~(\Wf)\footnote{$^\dagger$}{In deriving Eq.~(5.11)
we neglected a diagram where the $Z$ boson interacts with the $\pi^-$
before it is absorbed by the nucleon. We also neglected $\Delta$ degrees
of freedom and made no attempt to modify the $\pi$ properties for medium
effects. These issues are discussed in Migdal \etal~\qref\Pioncondensate\
in the context of a
degenerate nuclear medium. Although there is no real justification for
neglecting the same effects here, it does reduce the complexity of the result
and allows a simple comparison to other work\qref\Turner.}
$$\eqalign{
  \SPI = 16 \pi^2 \alpha_\pi
  \int&\frac{d^3\p}{(2\pi)^3 N_B} f_p(\p)\,[1-f_n(\p)]\,
  \int\frac{d^3\kkpi}{(2\pi)^3}\,\frac{f_\pi(\kkpi)}{2\opi}\,
  \frac{\delta(\opi + \o)}{[(p + \kpi/2)\cdot\kpi]^2}\,\times\cr
\alignskip
\alignskip
  &\times\,\biggl\{\caa^2 \kpi^\mu \kpi^\nu
   - \cab^2 \left[ P^{\mu\nu} \abs{\kkpi}^2 + P^{\mu\alpha} P^{\nu\beta}
             \kpi^\alpha \kpi^\beta \right] \cr
\alignskip
  &\hskip1cm
  - \cva^2 \left[ P^{\mu\nu} \opi^2 -u^\mu u^\nu \abs{\kkpi}^2
        - (u^\mu P^{\nu\alpha} + u^\nu P^{\mu\alpha}) \kpi^\alpha \opi
          \right]\biggr\}\,,\cr}
\EQ\Zb
$$
where $P^{\mu\nu} \equiv g^{\mu\nu} - u^\mu u^\nu$.
The isospin couplings are defined by $C_{A,0} =\half(C_{A,p}-C_{A,n})$
and $C_{A,1} =\half(C_{A,p}+C_{A,n})$, and similarly for the vector couplings.
The isospin 1 vector coupling does not contribute in the non-relativistic
approximation. Also, we have dropped a term proportional to $\cab \cva$
which is formally of the same order of magnitude but vanishes when
averaged over neutrino directions since it behaves as
$(\k_1 \times \k_2) \cdot \kkpi$.

The term $[(p + \kpi/2)\cdot\kpi]^{2}$ in the denominator of \eref{\Zb} comes
from the propagator of the intermediate nucleon. It differs slightly between
the two amplitudes which contribute to \eref{\Zb}, but we dropped the neutrino
4-momentum from the nucleon propagator so that the two diagrams would be more
similar. In the non-relativistic approximation and for thermal pions, it is
equal to $(m_N \opi)^2$; then the integral over the nucleon phase space
gives~1. However, for a pion condensate we will have to be more careful as
this denominator can diverge.

Note that the tensor structure of $\SPI$ reduces to the general form
of~\eref{\Wg} if the pions are distributed isotropically. However, if the pion
condensate has a non-zero momentum, the covariant Lorentz structure will
involve the condensate momentum and so, the number of distinct response
functions increases.

As a first specific case we discuss thermal pions. In order to determine the
contribution $S_\pi(\o)$ to our general response function we contract $\SPI$
with $N^{\mu\nu}$, and integrate over neutrino angles.  After this, any
angular dependence on $\kkpi$ has disappeared so that the pion phase space
integration can now be done just over $\opi$. For non-degenerate nucleons the
result is,
$$
\eqalign{S_\pi(\o) = & \frac{4\alpha_\pi}{m_N^2}\, (\o^2-m_\pi^2)^{1/2}\,
  \left[ \caa^2\,\left(1-\frac{m_\pi^2}{2\o^2}\right) +\cab^2\,\left(1-
  \frac{m_\pi^2}{\o^2}\right)
    + \cva^2\,\left(2-\frac{m_\pi^2}{2\o^2}\right)
      \right]\cr
\alignskip
& \times\cases{X_pf_{\pi^-}(|\o|)&for $\o < -m_\pi$,\cr
 \noalign{\vskip5pt}
   0 &for $-m_\pi < \o < m_\pi$, \cr
  \noalign{\vskip5pt}
   X_n[1+f_{\pi^-}(|\o|)]&for $\o>m_\pi$,\cr} \,. \cr}
\EQ\Zd
$$
If one takes $m_\pi =0$ and $\cab = 0$, this agrees with Turner's
result~\qref\Turner.
Note that the term with $\o>m_\pi$ corresponds to $n$ in the
initial and $p\pi^-$ in the final state, \ie, to the creation of a pion.
In thermal and chemical equilibrium one must satisfy
$\mupi = \mu_n - \mu_p$ so that
$ X_p f_{\pi^-}(|\o|) = e^{-|\o|/T} X_n [1+f_{\pi^-}(|\o|)] $, and
detailed balance is satisfied.

Since $S_\pi \propto \Theta(\abs{\o} - m_\pi)$, it is an explicit
example of a  ``hard'' $S$ with all power above some
threshold, so it is interesting to compare the scattering and pair process
strength as a function of $m_\pi$. Evaluating
$\Gamma_i = \int_0^\infty F_i(\o) S_\pi(\o)\,d\o$,
with $F_i$ given in Eq.~(\Xh)
for $i = {\rm scat}$ or pair, and using \eref{\Zd} we find that
$\gampair(x_\pi) = \gamscat(x_\pi)$ for $x_\pi \equiv m_\pi/T =  3.9$.
At this point the scattering processes are suppressed by about a
factor of 6 from their maximum if we had ignored the pion mass, \ie,
$\gamscat(3.9) = 1/6\, \gamscat(0)$; but the pair processes which are
sensitive to higher values of $\o$ are essentially unchanged from their
$x_\pi = 0$ value.

Next, we compare the thermally averaged scattering rate to the
single-nucleon scattering rate $\Gamma_\scat^N$, \ie, the scattering rate
expected if there were no multiple-scattering suppression.
Using $m_\pi = 0$, which maximizes $S_\pi$, and approximating
$1+f_{\pi^-}\to 1$ for thermal pions we have
$S_\pi(\o)=(4\alpha_\pi/m_N^2)\,X_n C_\pi^2\,\omega$ where
$C_\pi^2 = \caa^2 + \cab^2 + 2 \cva^2$. Then,
$$
  \frac{\Gamma_\scat^\pi}{\Gamma^N_\scat} =
  \frac{10 \alpha_\pi T^2}{\pi m_N^2}
    \approx 0.17
    \left(\frac{T}{50\,\MeV}\right)^2
    \left(\frac{800\,\MeV}{m_N} \right)^2 \,,
  \EQ\Ze
$$
where we have set $X_n C_\pi^2/((\SS_V + 3\SS_A)/4)=1$. By using the
single nucleon result, we have probably overestimated the contribution to
$\Gamma_\scat$ from nucleons; however, any multiple scattering suppression
of $S_\NN$ should also affect $S_\pi$. Further, we have overestimated
the contibution from pions by taking $m_\pi = 0$. Therefore, the
interaction of thermal pions with nucleons yields only a modest
correction to the scattering rate for the conditions pertaining to a SN core.

Now turn to the possibility of a pion condensate which implies that
$\mupi = \mu_n -\mu_p = \opo$ where $\opo$ is the lowest energy value for the
$\pi^-$. The interaction with nucleons typically causes a
dispersion relation for pions where this minimum occurs for a non-zero
momentum $\kpo$ so that the 4-momentum describing the condensate is
$k_{\pi0} = (\opo, \kpo)$. The occupation numbers for the condensate are given
by $f_\pi(\kkpi) = (2\pi)^3 N_\pi \delta^3(\kkpi - \kpo)$, reducing the
$\pi^-$ phase space integration in $\SPI$ to $N_\pi/(2\opo)$ and leaving a
factor $\delta(\opo -\o)$.

Most authors use a condensate with $\kpo \ne 0$, based upon consideration of
cold, degenerate nuclear matter characteristic of an older neutron star.  It
is conceivable, however, that  thermal effects may modify the pion dispersion
relation so that $\o_\pi(\kkpi)$ has its minimum at $\kpo = 0$.  We will first
consider such a zero-momentum condensate.  The contraction
$\SPI N_{\mu\nu}$ is identical to that for thermal pions
except that we have to set both $\opi=\opo$ and $m_\pi = \opo$. Then we find
$$
 S_\pi(\o) = 16 \pi^2\alpha_\pi\,\left(\caa^2 + 3 \cva^2 \right)
 \frac{N_\pi}{\opo m_N^2}\,
 \Bigl[X_p\delta(\opo+\o)+ X_n\delta(\opo-\o)\Bigr]\,.
\EQ\Zf
$$
We have used the fact that $1 + f_\pi \simeq f_\pi$ for the condensate. Then
$X_p/X_n=e^{-\opo/T}$, and detailed balance is satisfied.
Curiously, the axial and vector charges have exchanged roles from the single
nucleon case, a fact which is related to the $\gamma_5$ in the pion coupling
to nucleons.

It is easy to compare scattering and pair absorption for l.h.~neutrinos by
comparing $F_\scat$ with $F_\pair$ in Eq.~(\Xh) at $x=\opo/T$.  The pion
contribution to the pair process exceeds that to scattering for $\opo>6.4\,T$.
Equally, we may compare the energy-loss rate in r.h.\ neutrinos from
$\pi^- p\to n\overline\nu_L\nu_R$ with that from
$\nu_L \pi^- p\to n\nu_R$ or $\nu_L n \to \pi^- p \nu_R$. Here, the comparison
is between
$\widetilde F_\scat$ with $\widetilde F_\pair$ of Eq.~(\Qe). Pair emission
exceeds spin-flip scattering for
$\opo\gapprox 5.5\,T$. Therefore, in both cases scattering is more important
than the pair processes because for $T\approx50\,\MeV$ a condensate
with $\opo\gapprox5\, T$ is highly unlikely.

Next, we compare the pion-induced scattering rate with the single-nucleon one.
In analogy with Eq.~(\Ze) we now find
$$
  \frac{\Gamma_\scat^\pi}{\Gamma^N_\scat} =
  \frac{8 \alpha_\pi N_\pi}{m_N^2 T}
  \,\frac{x^2+6x+12}{6x}\,e^{-x}
  \EQ\Zee
$$
with $x\equiv\opo/T$.
For example, taking the number density of pions to be $N_\pi = 0.1 N_B$,
$m_N=800\,\MeV$, $T = 50\,\MeV$, and $\opo = 100\,\MeV$ we get
$\Gamma_\scat^\pi/\Gamma^N_\scat\approx 0.2$. Even for small values of
$\opo$ it is difficult to get pion-induced scattering to dominate
over that from single nucleons. Further, as $\opo$ is decreased the result
will become more sensitive to the scheme used to include multiple-scattering
effects for soft processes. To conclude, a zero-momentum
condensate is unlikely to dominate over single nucleon scattering for neutrino
transport or the emission of r.h.\ neutrinos.

Finally, we consider finite momentum condensates.
As long as $(\opo,\kpo)$ is time-like, we may always go to a Lorentz
frame with $\kpo=0$ at the expense of exact isotropy of the nucleon and
neutrino distributions. If $\abs{\kpo}/\opo$ is not large, then in the new
frame the fluid momentum will be small, number densities will be of order
their values in the fluid rest frame, and we may still use the long-wavelength
approximation. Apart from an overall factor
of order unity involving $\opo$ and $\kpo$, there will be no dramatic changes
from the case of a zero-momentum condensate.

However, when the condensate dips below the light-cone so that $(\opo,\kpo)$
is space-like, $\opo^2<|\kpo|^2$, a different approach is required.
In this
case the denominator $[(p+\kpi/2)\cdot\kpi]^2$ in Eq.~(\Zb) can become
zero, corresponding
to on-shell intermediate nucleon states. Put differently, the process $p\pi^-
\leftrightarrow n$ is now possible without any other particles involved.
However, because the pions have one fixed momentum, we may consider a Lorentz
frame where $\opo'=0$. There, the pion condensate looks like a static pion
field with a fixed wave-number $\kpo'$. The nucleons, then, should be
described as Bloch waves in this periodic potential, \ie, the nucleon
quasi-particles will be certain superpositions of $n$ and $p$, involving
spatial Fourier components of typical nucleon momenta and of $\kpo'$.

The NC scattering of neutrinos from these quasi-particles
in the non-relativistic and long-wavelength limits should be very similar to
the scattering off quasi-free single nucleons. Therefore, in the
non-degenerate limit it should be given essentially by the number density of
baryons times a typical weak cross section on a nucleon. We do not expect an
anomalously enhanced scattering rate. (We note that in old neutron stars a
pion condensate leads to a strong enhancement because the pion momentum is
available to conserve momentum and thus, processes can go which otherwise are
suppressed by degeneracy effects. In our case, there are no barriers from
momentum conservation to overcome, and scattering can occur with full strength
anyway.)

To summarize, we arrive at two conclusions.  First, pions would always enhance
the scattering rates more than they would enhance pair processes, whether or
not they are in a condensate. For r.h.\ neutrino emission that was considered
in Ref.~\Turner\ this means that its effect would be most important for the
spin-flip scattering channel $\nu_L p\pi^-\to n \nu_R$. Second, however, we
find that typically the effect of pions will not yield an anomalous
enhancement of the scattering rate relative to the single nucleon case.
The overall uncertainty of the neutrino interaction rates at high densities
appears to be much larger than the uncertainty introduced by the question of
whether or not there is a pion condensate.

We must comment that our conclusions differ somewhat from
Mayle, \etal~\qref\Wilson, who find that a pion condensate can have a
dramatic effect, especially for the emission of r.h.~neutrinos. There are
two essential differences. First, they estimate the emission of r.h.~neutrinos
by multiplying the emissivity per thermal pion times the number of pions in the
condensate, which we believe significantly overestimates the emissivity.
For example, we can compare the emissivity, $Q$, per pion by using
Eqs.~(\Zd) or (\Zf) for $S_\pi$, for thermal pions or for a $\kpo = 0$
condensate, respectively. Plugging into Eq.~(\Qd) and using
$\widetilde F_{\rm pair}$, the ratio of the emissivity per
thermal $\pi$ divided by the emissivity per condensate $\pi$ is,
$$
\frac{(Q/N_\pi)_{\rm th}}{(Q/N_\pi)_{\rm cond}} =
   \frac{30 C^2_\pi}{(\caa^2 + 3 \cva^2)\, x_0^3\, (1 + e^{-x_0})}\,.
  \EQ\Zg
$$
Unless the condensate is ``hard'', use of the thermal pion formula
overestimates the emissivity by an order of magnitude; but, if
$\opo > T$ then Eq.~(\Zee) shows that the rates will be small
compared to those from single nucleons. Further, as argued above, for a
$\kpo \ne 0$ condensate we expect little enhancement over the single nucleon
emissivity, if one uses an appropriate set of Bloch states to
describe the nucleons.

We have conducted our entire discussion of the pion-induced effects as if we
were in a dilute medium where different contributions to the emissivities add
linearly. The pion interactions, however, are just another contribution to the
nucleon spin fluctuations and so, they essentially add to the width of
$S_A(\omega)$ without adding to the overall strength, at least
as long as we may treat the normalization in Eq.~(\SSf) as an upper bound.
If we take our conclusions concerning $S_\NN$ and $S'_\NN$ as a guide,
it is quite possible that a large number of pions could actually
have the effect of {\it decreasing\/} the effective emissivities of axions or
r.h.\ neutrinos.

The second distinction is that
Mayle, \etal\ \qref\Wilson\ stress that the presence of charge in a pion
condensate will reduce the electron degeneracy, which in turn results in the
release of entropy and heating of the core material. The resulting increase in
temperature increases all emission rates. In this picture, the presence of
pions
affects the emission rates indirectly through their impact on the equation of
state. This aspect of Mayle \etal's work is precisely in the spirit of what we
advocate: When adding novel physics to the model one must be careful to include
all ramifications, not just focusing on a single aspect
of the new phenomena. In the present context, if one adds a
pion condensate to the model to increase the emission of r.h.~neutrinos
then one must also allow for a change in the transport properties of the
ordinary l.h.~neutrinos.




\chapter{Discussion and Summary}

\noindent
We have studied weak NC processes in the environment of the newly born neutron
star in a SN core. They are crucial for understanding the cooling of the core
for the first tens of seconds after collapse, the time during
which the neutrino flux from a galactic SN would be observable in a large
underground detector. The most important application is the role NCs play in
ordinary neutrino transport of lepton number and heat. In addition, they also
couple weakly to a variety of hypothetical particles,
which would provide for an anomalous loss of heat from the core by particle
emission. Constraints on exotic particle properties can be derived
from a comparison between their calculated effect on the neutrino flux from
the ``surface'' of the star with the SN~1987A observations.

Understanding these transport and emission processes is made difficult because
of the breakdown of perturbation theory, the usual tool for studying particle
interactions. Nonetheless, it is possible to relate different aspects of a
particular problem, \eg\ the transport of
lepton number or energy,
to a common medium structure function, $S(\o)$, where $\o$
is the energy transfer to the medium. The rate for any weak process can
then be put into the form $\int d\o\,F(\o) S(\o)$, where $F$ is specific
to that process. The difficulty with perturbation theory is confined
to the function $S(\o)$, so that once a particular $S(\o)$ has been
determined, the rates for all related processes can be calculated in a
consistent fashion. Of course, different processes have $F$'s which weigh the
frequency dependence of $S$ in different ways, so there remains some
sensitivity to the shape of $S(\o)$.  Within this framework we have compared
neutrino scattering and pair absorption and emission as they apply to the
transport of ordinary neutrinos.  We have also studied how the same
functions $S(\o)$ relate to the emission of r.h.~neutrinos and axions.

In the limit that the medium is dominated by non-relativistic nucleons,
the medium response consists of the sum of vector and
axial vector response functions, $S_V(\o)$ and $S_A(\o)$ respectively.
In the context of perturbative calculations of $S$, we have discussed two
contributions: one from quasi-free nucleons $S_0\delta(\o)$, and one from
interactions in the presence of two nucleons $S_\NN(\o)$. In the
non-relativistic long-wavelength limit the former is a delta function,
whereas the latter naively has a soft divergence $S_\NN(\o) \sim \o^{-2}$.
In \S~III we argued that, based on physical arguments and explicit
calculation, $S_\NN(\o)$ contributes only to $S_A$, \ie\
the vector response is well approximated by quasi free nucleons.

The axial response on the other hand requires a full calculation of the
spin fluctuations in the medium, and is not well determined.
Nucleon-nucleon interactions randomize the nuclear spin at a rate
comparable to the collision frequency, which has a two-fold effect.
They soften $S_A\delta(\o)$ to something like a Lorentzian with a
width of about the nucleon-nucleon interaction rate $\gamint$ and,
by the same token, they regulate the divergence of $S_\NN(\o)$ at
$\o=0$. It is possible to choose the width of the Lorentzian so
that it reproduces the quasi-free nucleon $\delta$-function result in
the limit of a very dilute medium, while at the same time for large values
of $\o$ the bremsstrahlung result for $S_\NN(\o)$ is obtained.
Although this formulation is appealing (it replaces the one
and two nucleon results by a single function), if extended into
the regime of large collision rates it implies a very strong suppression
to the axial response function and, as discussed in \S~III,
the internal consistency of the approach may be questioned, as well as
its justification based on a comparison with laboratory data.

We conclude that there is no unique way to extrapolate the $S_\NN$ to high
densities.
One can obtain anything from an enhanced to a much
suppressed neutrino scattering rate.
We stress that this inadequacy of perturbation theory cannot be ignored.  We
are
interested in the evolution of
a neutron star for, say, the first ten seconds after the collapse of the
progenitor star. During this phase, all numerical models show maximum
temperatures around $T=50\,\MeV$.  Even though large parts of the core may be
cooler, the neutrino opacity increases with temperature, and so the hot
regions should act as a bottleneck to neutrino transport.  There, the nuclear
medium is on the verge of degeneracy, but still essentially non-degenerate. If
the medium were sufficiently dilute, then even for strong interactions one
could proceed perturbatively with some confidence, while if the degeneracy
were much higher only nucleons near their Fermi surfaces participate, in
effect rendering the medium more dilute.  In the middle, perturbation theory
fares worst.  Still, because one can calculate reasonable answers both for
very high and for very low degeneracies, we find it difficult to believe that
the neutrino scattering rate should be {\it very\/} different from its naive
lowest-order result in the intermediate range. In particular, an anomalously
{\it enhanced\/} scattering rate seems rather unlikely,
especially as it would require a violation of the upper limit in
Eq.~(\SSf), which is based on the presumption that the different
momentum, spin, and isospin parts of the nucleonic wave function are
uncorrelated.

We speculate that the overall shape of $S(\o)$ given by the low-density
approximation will not be radically modified at high densities. In this case
scattering rates definitely exceed the rates for pair processes. This
conclusion applies to the transport of l.h.~neutrinos as well as to the
emission of r.h.~ones. Moreover, extreme modifications to the shape
of $S_A$ seem likely to suppress the total interaction rate for both
scattering and pair processes. Since the vector response is essentially
unmodified and contributes only to scattering, it seems unlikely that any
reasonable form for $S_A$ will make pair processes dominate over
scattering. This conclusion holds equally for the scattering of
l.h.~neutrinos and the emission of r.h.~neutrinos.

We caution, however, that there seems to be a trend that as the medium becomes
less dilute
the response function gets suppressed for low $\o$ while it increases on the
wings. We have assumed that for $\o$ above a typical nucleon-nucleon
interaction
rate, $S$ is still reasonably represented by a perturbative result.  If this
were not the case, the true shape of $S(\o)$
could be so ``hard'' that pair processes could be important for neutrino
transport after all. Equally, if the medium had strongly excited collective
modes, their decay into neutrino pairs could be stronger than their
contribution to scattering, depending on the nature of their dispersion
relation. In this light we note that Iwamoto and Pethick~\qref{\IwamotoP} have
discussed the importance of sound waves with a strong NC
coupling, while Haensel and Jerzak~\qref{\HaenselJ} have shown that in
degenerate matter quasi-free nucleons still seem to make the dominant
contribution. Further, in the explicit case of a time-like $\pi^-$ condensate
we find that pair emission does not dominate.

Given the uncertainty in the basic transport rates, how is one to understand
the good agreement between the SN~1987A neutrino signature with theoretical
expectations? One answer is that other parameters may have been adjusted to
achieve good agreement with the SN~1987A neutrino signal, hiding the
sensitivity to the neutrino transport. Another possibility is that somehow the
cooling curve is rather insensitive to the details of neutrino transport.
Or, perhaps the transport in the numerical models, just by chance, is close to
the real thing. For example, if the axial response were eliminated
entirely, the vector resoonse would still contribute $\approx 1/5$ of the
naive single nucleon scattering rate. Perhaps even a fifth of the
standard scattering rates is sufficient to reproduce the observed cooling
timescale of SN~1987A. In this light, we feel it is time
to undertake a systematic survey of how variations in the
neutrino diffusion affect the cooling of the core on timescales
from a second to a minute. This certainly seems important
in light of the efforts to prepare experiments in advance
of the next galactic supernova neutrino burst.

We have also commented on the derivation of bounds on exotic particle
interactions. In such calculations, the non-standard processes should be
treated on the same footing with the ordinary neutrino transport, \ie, all
NC type processes should be based on the same function $S$.
Any regulation applied to keep the ordinary scattering rates bounded must be
applied consistently to exotic emission rates also, which essentially has the
effect of decreasing them. Further, new processes proposed to increase
emission rates must be studied for their effects on neutrino transport as
well. In the end, only calculations where the transport and emission are
performed with the same approximations for $S(\o)$ may be used for
quantitatively precise constraints, and even then only after allowing for
different choices of $S(\o)$ and variation of all other unknown parameters.

Finally, we have evaluated the contribution to the response function
from the interaction of single pions with nucleons, $S_\pi(\o)$.
Contrary to previous speculations in the literature, we
find no indication for an anomalously large contribution by this
latter process for the conditions relevant for the cooling of a SN core.

\nochap{Acknowledgements}
\noindent We thank S.~Blinnikov, T.~Janka, W.~Hillebrandt, J. Engel,
and S.~Pittel for helpful discussions.
D.~S.\ acknowledges financial support from the Max-Planck-Institut
f\"ur Physik, DOE grant DE-AC02-78ER05007, and
the University of Delaware Research Foundation.

\newpage




\appendix{Effective Nucleon Weak Charges}

\noindent
If the nucleon weak current is written in the
form Eq.~(\Wee) the vector charge is given by
$C_V=\tau_3-4\sin^2\theta_W Q$, where $\tau_3=+1$ for protons and $-1$
for neutrons and $Q$ is the nucleon electric charge in units of $e$.
Therefore,
$$C_V^n=-1\hbox{\quad and\quad} C_V^p=0.07\,,\EQ\AAa$$
where we have used $\sin^2\theta_W=0.2325$ for the weak mixing angle.

There remain problems, however, with the value of the axial charge. If
axial currents were conserved we would have $C_A=\tau_3$, but axial
charge conservation is known to be violated by the strong interactions,
leading to a charged-current axial coupling of bare nucleons of
$C_{A,CC,0}=1.26$. In large nuclei this value is suppressed somewhat,
and the commonly used value~\qref\axnucl\ for nuclear matter
calculations
is $C_{A,CC}=1.0$. Based on the naive quark model it was common practice
until recently to take the NC axial charges to be isovector
in character and so, $C_{A,NC}=1.0\,\tau_3$ was a reasonable choice.  It
is now realized, however, that the neutral axial current has an
isoscalar piece as well, which is associated with the polarization of
the strange quark sea of nucleons.

The axial charge of nucleon $i$ can be written as a sum over the
contributions from different quarks,
$C_A^i=\Delta u^i-\Delta d^i-\Delta s^i$.
The contributions from $u$ and $d$ quarks include both valence and sea
contributions, whereas the $s$ quark only has a sea contribution.  It is
expected that the heavy quarks do not contribute significantly.  Isospin
symmetry dictates that $\Delta u^n = \Delta d^p$ and $\Delta d^n =
\Delta u^p$. Charged current processes determine $\Delta u^p - \Delta
d^p = 1.26$ for bare nucleons. Scattering experiments of polarized muons
on polarized hadronic targets by the EMC~\qref\EMCa\ and SMC~\qref\SMC\ groups
at CERN, and by polarized electrons at SLAC (E142~\qref\Eslac) give a
range of values for  $\Delta s^p$ between 0 and $-0.2$. In a review of
the phenomenolgy and theory, Ellis and Karliner~\qref\EllisK\ suggest
$\Delta s = -0.11 \pm 0.04$,
leading to $C_{A,0}^p = 1.37$ and $C_{A,0}^n  = -1.15$ in
vacuum. For nuclear matter we find
$$C_A^n=-0.91 \quad{\rm and}\quad C_A^p=1.09
  \EQ\AAb$$
if $\Delta u$, $\Delta d$, and $\Delta s$ are suppressed by an equal
factor $1/1.26$. We do not give uncertainties because of the many
possible sources of systematic errors, both in the interpretation of the
laboratory results and in the extrapolation to nuclear matter.

For the scattering of neutrinos on quasi-free nucleons the relevant
combination of coupling constants is
$\overline{C^2}=\quarter(C_V^2+3C_A^2)$ for which we find
$$\overline{C^2}=\cases{0.89 &for protons (1.41 in vacuum),\cr
                        0.87 &for neutrons (1.24 in vaccum).\cr}
  \EQ\AAc$$
Therefore, the scattering rate is suppressed by about a factor of 0.63
for protons and 0.70 for neutrons relative to their vacuum values.


\newpage

\refout

\newpage

\centerline{FIGURE CAPTIONS}
\bigskip
\bigskip

\centerline{FIGURE 1}
\medskip
\noindent{The functions $F_\scat$ and $F_\pair$ defined in Eqs.~(\Xg\ and
\Xgg), and the analytic approximations defined in Eq.~(\Xh).
The results are normalized to $F_0 \equiv \GF^2 N_B T^2$}.

\bigskip
\bigskip

\centerline{FIGURE 2}
\medskip
\noindent{Comparison of scattering and pair absorption rates, Eq.~(\Xf),
calculated using the $F$'s in Eq.~(\Xh), for different choices of $S(\o)$.
Horizontal lines are single-nucleon scattering rates.
Dotted curves use $S_\NN$ from Eq.~(\Yc) with a lower cutoff $\omin$ while the
dashed curves are for $S'_\NN$ from Eq.~(\Yd), also cut off below $\omin$.
The parameters used are $\rho = \rho_0 = 3 \times 10^{14}$ gm cm$^{-3}$,
$T = 50\,\MeV$, $X_n = 0.7$, $m_N = 800\,\MeV$, and the vector
and axial charges given in Eqs.~(\AAa) and (\AAb), respectively.}

\bigskip
\bigskip

\centerline{FIGURE 3}
\medskip
\noindent{Comparison of scattering and pair emission contributions to the
emission of r.h.~neutrinos from Eqs.~(\Qd) and (\Qe), for the same choices of
$S(\o)$ and parameters as in Fig.~\fGamma.}


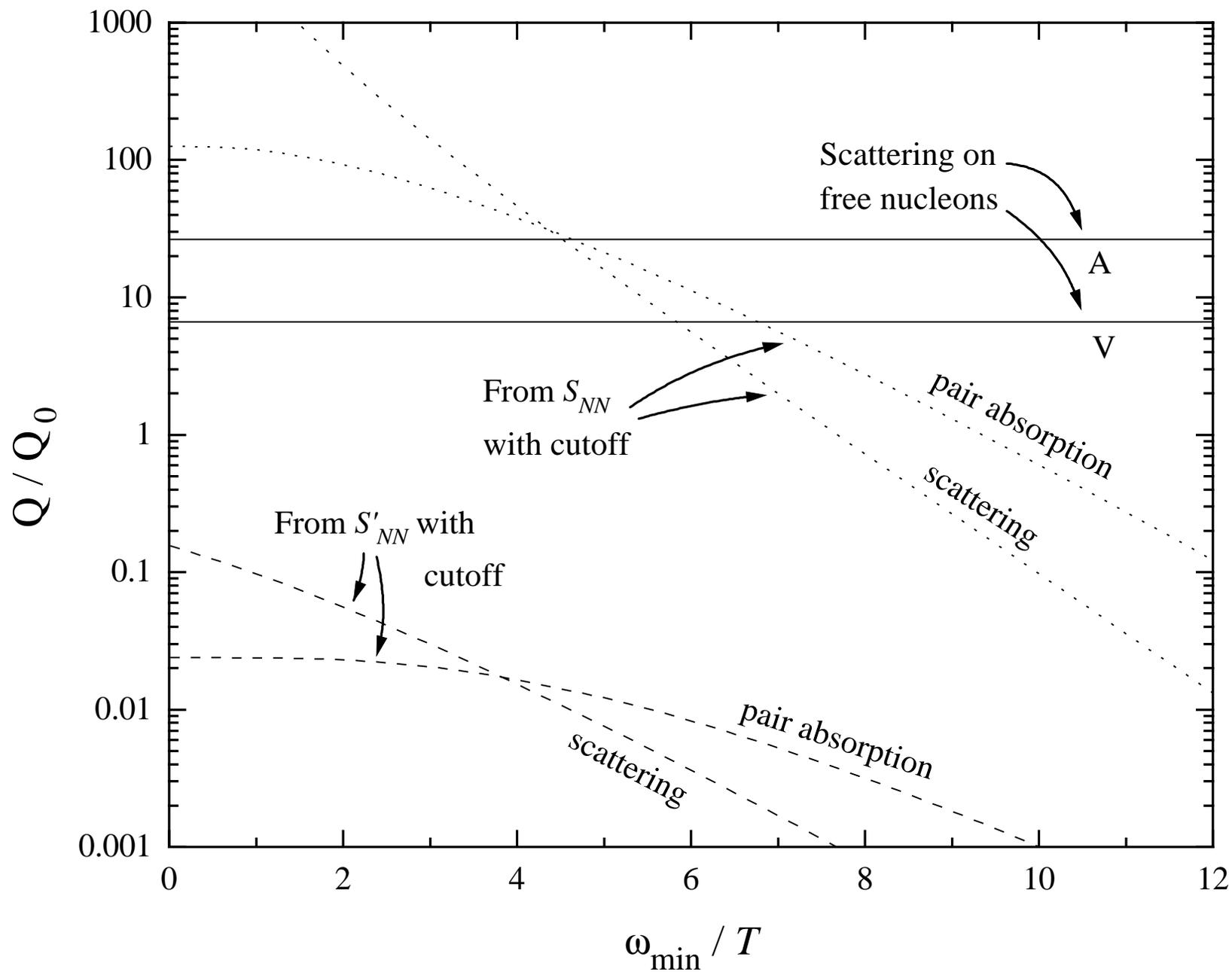
\bye